\documentclass{aa}

\usepackage{longtable}
\usepackage[varg]{txfonts}
\usepackage{graphicx}
\usepackage{hyperref}
\usepackage{multirow}
\usepackage{supertabular,booktabs}
\usepackage{float, caption, booktabs, cellspace}
\usepackage{xcolor}
\usepackage{lscape}

\begin{document} 

\title{\textbf{Cometary dust collected by MIDAS on board Rosetta\thanks{Tables \ref{table:target exposure history} and \ref{table:Best representation of 1857 particles} are only available in electronic form at the CDS via anonymous ftp to \url{cdsarc.cds.unistra.fr} (\url{130.79.128.5}) or via \url{https://cdsarc.cds.unistra.fr/cgi-bin/qcat?J/A+A/}}}}
\subtitle{I. Dust particle catalog and statistics}
\titlerunning{Dust of comet 67P - MIDAS particle catalog and statistics}

\author{M.~Kim\inst{\ref{inst1}, \ref{inst2}, \ref{inst3}} \and
T.~Mannel\inst{\ref{inst1}} \and
P.~D.~Boakes\inst{\ref{inst1}} \and
M.~S.~Bentley\inst{\ref{inst4}} \and
A.~Longobardo\inst{\ref{inst5}} \and
H.~Jeszenszky\inst{\ref{inst1}} \and
R. Moissl\inst{\ref{inst6}} \and the MIDAS team}

\institute{{Space Research Institute of the Austrian Academy of Sciences, Schmiedlstrasse 6, 8042 Graz, Austria\\ \email{minjae.k.kim@warwick.ac.uk}}\label{inst1}
\and
{Department of Physics, University of Warwick, Gibbet Hill Road, Coventry CV4 7AL, UK}\label{inst2}
\and
{Centre for Exoplanets and Habitability, University of Warwick, Gibbet Hill Road, Coventry CV4 7AL, UK}\label{inst3}
\and
{European Space Astronomy Centre, Camino Bajo del Castillo, s/n., Urb. Villafranca del Castillo, 28692 Villanueva de la Cañada, Madrid, Spain}\label{inst4}
\and
{Istituto Nazionale di Astrofisica, Istituto di Astrofisica e Planetologia Spaziali, via Fosso del Cavaliere 100, I-00133 Rome, Italy}\label{inst5}
\and 
{Scientific Support Office, Directorate of Science, European Space Research and Technology Centre, 2201 AZ Noordwijk, The Netherlands}\label{inst6}}

\date{Received 21 October 2022 / Accepted 15 February 2023}


\abstract
{The Micro-Imaging Dust Analysis System (MIDAS) atomic force microscope (AFM) on board the Rosetta comet orbiter has been dedicated to the collection and 3D topographical investigation of cometary dust in the size range of a few hundreds of nanometers to tens of micrometers with a resolution down to a few nanometers.}
{We aim to catalog all dust particles collected and analyzed by MIDAS, together with their main statistical properties such as size, height, basic shape descriptors, and collection time. Furthermore, we aim to present the scientific results that can be extracted from the catalog, such as the size distribution and statistical characteristics of cometary dust particles.} 
{Through a careful re-analysis of MIDAS AFM images, we make a significant update and improvement to the existing MIDAS particle catalog, resulting in the addition of more particles and newly developed shape descriptors. The final product is a comprehensive list of all possible cometary dust particles detected by MIDAS. The catalog documents all images of identified dust particles and includes a variety of derived information tabulated one record per particle. Furthermore, the best image of each particle was chosen for subsequent studies. Finally, we created dust coverage maps and clustering maps of the MIDAS collection targets and traced any possible fragmentation of collected particles with a detailed algorithm.}
{The revised MIDAS catalog includes 3523 MIDAS particles in total, where 1857 particles are expected to be usable for further analysis (418 scans of particles before perihelion + 1439 scans of particles after perihelion, both after the removal of duplicates), ranging from about 40~nm to about 8~$\mu$m in size. The mean value of the equivalent radius derived from the 2D projection of the particles is 0.91~$\pm$~0.79~$\mu$m. A slightly improved equivalent radius based on the particle's volume  coincides in the range of uncertainties with a value of 0.56~$\pm$~0.45~$\mu$m. We note that those sizes and all following MIDAS particle size distributions are expected to be influenced by the fragmentation of MIDAS particles upon impact on the collection targets. Furthermore, fitting the slope of the MIDAS particle size distribution with a power law of ${a \cdot r}^{\rm{\,b}}$ yields an index $\rm{\,b}$ of $\sim$ -1.67 to -1.88. Lastly, based on the created dust coverage maps and clustering maps of the MIDAS collection targets, we determined the particle fragmentation ratio of 4.09 for nominal activity and 11.8 for the outburst, which underlines that parent particles with faster impact velocity are more likely to be fragmented during dust collection.}
{}



\maketitle

    \section{Introduction}

Since comets are small bodies with orbits generally far from the Sun, they store material that remains almost entirely unaffected by gravitational pressure and solar irradiation. Consequently, comets are thought to have preserved dust from the very beginning of the Solar System's formation. In particular, the smallest cometary dust grains are of great importance as they represent the building blocks of planetesimals and planets and, thus, they contain information on their earliest growth processes thought to start with collisional aggregation of the smallest dust grains (\citealp{tielens_dust_2005, li_dust_2003, weidenschilling_formation_1993, blum_growth_2008}). 

The microscopic properties of the smallest dust particles have been studied in a variety of ways, for instance, samples returned from comet 81P/Wild 2 (\citealp{brownlee_comet_2006}), chondritic porous interplanetary dust particles (CP-IDPs) collected in the stratosphere (\citealp{flynn_organic_2013}), and Antarctic (\citealp{noguchi_dust_2015}), as well as ultracarbonaceous Antarctic Micrometeorites (\citealp{duprat_ucamms_2010}). However, these investigations either suffer from significant dust alteration between the time of release from the comet and their analysis or their provenance is unknown.
Although it is generally difficult to obtain structural information of pristine cometary dust particles, a unique opportunity to sample the dust and gas environment of the inner coma of a comet, namely, 67P/Churyumov-Gerasimenko (hereafter 67P), was offered by the Rosetta mission between 2014 and 2016. In particular, dust particles were collected with low relative velocities (roughly between some cm~s$^{\rm -1}$ and tens of m~s$^{\rm -1}$; \citealp{dellacorte_giada_2015, fulle_density_2015}) enabling dust particles to be collected with only a small degree of alteration after their release. One of three in situ dust instruments on board, Micro-Imaging Dust Analysis System (MIDAS;~\citealp{riedler_midas_2007, bentley_lessons_2016}) was the first atomic force microscope (AFM) that was launched into space to collect and investigate the properties of cometary dust. It collected particles with sizes from hundreds of nanometers to tens of micrometers and recorded their 3D topographic and textural information and statistical parameters (\citealp{bentley_lessons_2016, bentley_morphology_2016}). We note that AFMs have been developed for planetary science applications to study extraterrestrial samples such as Martian soil particles (Phoenix/NASA; \citealp{Vijendran2007, Smith_Martian_soil}), lunar samples (Apollo 11/NASA; \citealp{Hammond_2007}), and atmospheric aerosol particles (\citealp{Lee_Tivanski_2021}) as well.

The  findings of MIDAS presented in this work are set in relation to those of the other dust analysis instruments on board Rosetta: COmetary Secondary Ion Mass Analyzer (COSIMA; \citealp{langevin_typology_2016}), which was dedicated to the chemical analysis of cometary particles $\sim$ 10 ${\mu}m$ to submillimeter in size and delivered optical microscope images with resolutions down to about 10 micrometers, and Grain Impact Analyser and Dust Accumulator (GIADA;  \citealp{colangeli_GIADA_2007}), which derived the number, mass, momentum, and velocity of dust particles from comet 67P.\newline

\noindent The goal of the current study of MIDAS data is to present the revised MIDAS particle catalog, together with the main statistics and scientific results{\footnote{The definition of language used in this study can be found in Appendix~\ref{sec:terminology}.}} that can be extracted. This paper is organized as follows: we provide a description of the MIDAS instrument and the mechanism of dust collection in Section~\ref{sec:methods}. Furthermore, we discuss a sophisticated clustering algorithm to understand the fragmentation of dust particles during collection. In Section~\ref{sec:results}, we give a general description of the MIDAS particle catalog and its main statistics. Based on the results, we show the size distributions of MIDAS particles and fragments. Furthermore, we present dust coverage maps and clustering maps showing the distribution of fragments and parent particles. We summarize our findings in Section~\ref{sec:conclusions}.

    \section{Methods}\label{sec:methods}
        
\subsection{Description of the MIDAS AFM, its dust collection and imaging}\label{MIDAS_AFM}

\begin{figure*}
\includegraphics[width=18cm, height=6cm]{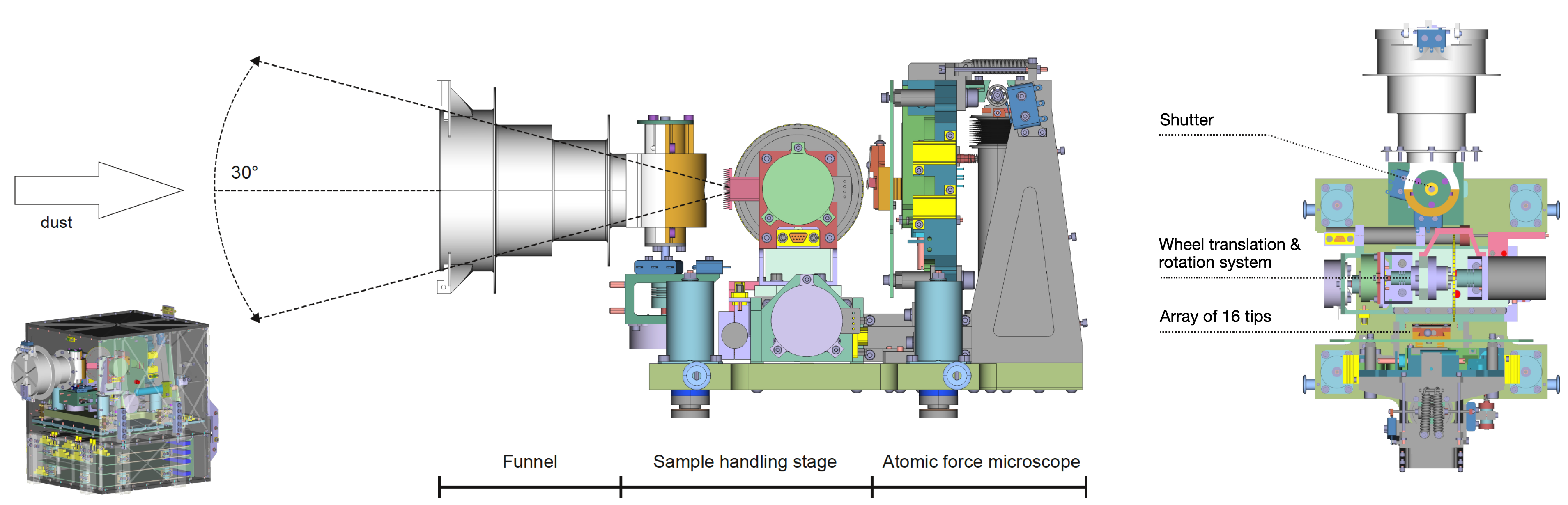}
\caption{MIDAS instrument parameters in detail (based on \citealp{ESLAB_Bentley_2016}).}
\label{fig:MIDAS_instrument}
\end{figure*}

MIDAS was the first in situ AFM launched into space. MIDAS had a passive dust intake system and collector as presented in Figure~\ref{fig:MIDAS_instrument}. It consisted of a funnel leading the particles through a shutter that controlled the particle flux and thus the exposure time onto the collection targets (sized 1.4 mm x 2.4 mm with an area of about 3.5 mm$^{2}$) mounted on the circumference of a sample wheel (\citealp{bentley_lessons_2016}). The targets could be transported to an AFM stage equipped with 16 small, very sharp tips that rastered the surface of the targets and the collected dust particles. The amplitude-modulated AFM obtained resolutions down to a few nanometers per pixel (\citealp{ESLAB_Bentley_2016}).

The resulting data product is a 2D array where every entry stands for the relative height measurement at a certain location by the change in amplitude of the AFM oscillating cantilever. We note that the assessed 3D topographies are the convolution of the structure of the particle and the tip shape, where the smallest influence can be expected for particles with rims (i.e., vertical topography with a gradient shallower than the tip opening angle), much flatter than the steepness of the tip's apex, as for all AFM measurements. As a result, MIDAS acquired 3D images of cometary dust that can be used to analyze the shape, volume, topographic structure and morphology, and other parameters of individual particles.

Dust particles entering MIDAS had to pass a funnel with a 30-degree field of view that narrowed in steps (see Figure \ref{fig:MIDAS_instrument}) and ended in a shutter with a hole sized like a MIDAS target (1.4 mm $\times$ 2.4 mm). Thus, no particles larger than a millimeter could enter MIDAS and only small particles are assumed to finally collide with the collection target and thus be collected. Additionally, the maximal scan size was about $\sim$~80~$\mu{\rm{m}}$ $\times$ 80 $\mu{\rm{m}}$, such that the largest detected particle was several tens of micrometers in size. We note that MIDAS targets are fixed relative to the spacecraft's movement and orientation because the Rosetta spacecraft kept its orientation relative to the spacecraft velocity vector with an operational requirement of the OSIRIS cameras for the Sun illumination to always be coming from the same direction.

In total, 61 targets were available for dust collection, with 3 pre-mounted targets used for in-flight calibration. After the arrival of the Rosetta spacecraft at comet 67P, 8 targets were exposed and scanned for dust collection (15 targets were exposed in total from the start of the mission), but only 4 targets (i.e., targets 11, 13, 14, and 15\footnote{Note: Previous studies (e.g., \citealp{Longobardo_2022_MNRAS, Kim_Mannel_2022EGU}) have adopted different indexing methods, which comes from the number of telemetries counted from 1 to 64. However, the MIDAS particle catalog and browse images count the targets from 0 to 63. Thus, target numbers 10, 12, 13, and 14 in the previous studies correspond to target numbers 11, 13, 14, and 15, respectively, in this study.}) were found to have detectable dust deposits of cometary origin. The detailed list of target exposure history can be found in Table \ref{table:target exposure history} of Appendix~ \ref{appendix: Target exposure history}. Other targets (e.g., target 07) were found to show features that are similar to cometary dust but are probably not of cometary origin. The most likely reason is contaminant particles on the target being imaged with a tip that has cometary dust stuck to it (due to the previous scanning on a target with cometary dust). Tip sample convolution then leads to features similar looking to cometary dust particles despite no cometary dust being collected on the target. Furthermore, each target collected dust in a different time period with differing arriving particle speeds $v$ (\citealp{Longobardo_2020a_Merging_data}). For example, targets 11 and 13 collected tiny dust particles in three periods before perihelion (Sep. – Nov. 2014, Dec. 2014 – Feb. 2015, and Feb. – Mar. 2015, respectively) with arriving particle speed $v$ about 3m~s$^{\rm -1}$; whereas the dust particles collected on target 14 were released in an outburst after perihelion (Feb. 2016), with an arriving particle speed, $v,$ about 7m~s$^{\rm -1}$~(\citealp{Gruen_2016_outburst, Longobardo_2022_MNRAS}). The different periods again correspond to different activities of the comet as well, for instance, nominal activity at pre-perihelion and the February 16, 2016 outburst at post-perihelion~(\citealp{Gruen_2016_outburst}). 

We note that these velocities are measured by GIADA (\citealp{Rotundi_2015}), whose detected particle size ranges from about 50 $\mu{\rm{m}}$ to several millimeters (up to two orders of magnitude larger than MIDAS' particles). However, the constant dust flux ratio measured by GIADA regardless of periods (nominal activity and outburst) and the results from the activity model (\citealp{Longobardo_2022_MNRAS}) concluded that the particles detected by MIDAS are fragments of hundreds-micron- to mm-sized particles detected by GIADA. Thus, we assume that the speed of MIDAS parent particles is similar to those detected by GIADA. We also note that individual small particles may have arrived with much higher velocities (\citealp{Agarwal2007, dellacorte_giada_2015, Merouane_COSIMA_size_D_2016, hilchenbach_mechanical_2017}). However, small individual particles are rarely found in the MIDAS dataset, thus, we assume that the majority of particles arrived with velocities of several meters per second as measured by GIADA.

A typical scientific sequence of MIDAS was started with a prescanning of an area of a target, exposing it to the dust flux of the comet, and rescanning of the same area with low resolution. If a dust particle was detected, it was then scanned with optimized parameters (\citealp{bentley_lessons_2016, ACLR2018}). To perform the grain identification, analysis, and calibration, the open-source package Gwyddion\footnote[2]{\url{http://gwyddion.net}} was used throughout this study. In particular, we drew the mask by hand, judging by eyes, but cross-checked it iteratively (\citealp{bentley_morphology_2016}) More detailed information about the MIDAS instrument can be found in previous MIDAS publications (\citealp{riedler_midas_2007, bentley_lessons_2016}).

\subsection{Description of the MIDAS clustering algorithm:\ Identification of fragments with common origins}\label{sec:clustering algorithm}

When a dust particle entered the MIDAS instrument, it collided with the collection target with a certain speed, so the particles could be fragmented upon impact. A fragmentation resulted in a characteristic pattern of dust distribution on the target, producing so-called clusters, where all particles in the same cluster are expected to have a common origin (i.e., coming from the same parent particle). In particular, the initial particle size (and structure) and particle velocity (\citealp{Ellerbroek_labstudy_2017}), as well as the composition can strongly determine how far the fragments of clusters were spread.

We aim to determine which MIDAS individual particles are fragments of the same parent particle. In principle, the proximity (i.e., the distance between the data points to each other; centroid to fragments) of dust particles was taken into account in this work to identify all fragments of one parent particle. This can be done by hand, but to avoid bias and facilitate repeatability, a clustering algorithm was considered. 

Given a set of data points, a clustering algorithm classifies each data point into a specific group. Among a broad variety of available classification algorithms, it is critical to choose the most suited one for the given data set. In particular, the clustering result should find cluster centers that minimize conditional variance (i.e., an appropriate representation of data) and be repeatable and consistent. Furthermore, a non-parametric algorithm (i.e., no assumption of any prior shape of the final cluster such as spherical, elliptical, or similar) was used. Thus, the clustering result is not sensitive to the initial parameters and is not dependent on too many parameters. Lastly,  the clustering result should not be sensitive to outliers; outliers have to be considered as single particles forming their own cluster, not one of the components in a wrong cluster. Based on these requirements, we chose the mean shift-clustering method (\citealp{mean_shift_clustering}) among several clustering algorithms and approaches. In particular, mean shift clustering does not need the number of clusters to be specified in advance because the number of clusters will be determined by the algorithm with respect to data unlike in most other algorithms, such as $k$-means clustering or the Gaussian mixture model (\citealp{k_mean_clustering, Numerical_Recipes_3rd_Edition}). 
In the mean shift-clustering method, given $n$ data points $ x_{\,\rm{i}}$, on a d-dimensional space $ R^{\,\rm{d}}$, and  (as the only free parameter ) the window radius (called the bandwidth), $h$, the multivariate kernel density, $f(x),$ estimate obtained with kernel K($x$) can be described as:
\begin{equation}
f(x) = \frac{1}{n\cdot h^{\,\rm{d}}} \,\Sigma\,\mathrm{K}\,\bigg(\frac{x - x_{\,\rm{i}}}{h}\bigg).
\label{mean_clustering}
\end{equation} 
        
Based on the kernel density estimation, the mean shift algorithm assigns the data points to the clusters iteratively by shifting points (a so-called sliding-window-based algorithm) towards the highest density of data points (i.e., cluster centroid). The choice of a sensible bandwidth, $h,$ is described in Section~\ref{sec:results_Fragmentation} and Table~\ref{table:band_width_value_cluster}. The resulting clusters and implications for MIDAS dust collection and comparison to other data are discussed in Section~\ref{sec:results_Fragmentation} and Appendix~\ref{appendix: band_width}.
Visualizations of the final clustering results can be found in Appendix \ref{appendix_dust_clustering_maps}.

        
\section{Results and discussion}\label{sec:results}

\subsection{MIDAS particle catalog}\label{sec:MIDAS_catalog}
\subsubsection{Description of the MIDAS particle catalog}\label{sec:Description_of_the_MIDAS_particle_catalog}

In the MIDAS particle catalog, all particles detected by MIDAS together with all available information about them are documented and displayed as a table. Each particle is described in one line where its (meta) data are collected in columns, for instance, images of single particles and various statistical evaluations of the particles according to size, volume, and basic shape descriptors. In particular, each cometary dust identification in every MIDAS AFM scan is masked using the open-source Gwyddion AFM software (\citealp{Gwyddion_paper}) and a link to the particle mask file in the catalog is provided along with the 3D topographic data of each particle. Additionally, several flags highlighting their origin are given, the quality and/or confidence of the identification and image and scan quality, as well as additional parameters describing the particle and image properties. In the present study, we improve on the first public version of the MIDAS catalog (MIDAS particle catalog V5\footnote[3]{ESA planetary science archive of MIDAS/Rosetta (PSA 6.2.3), dataset identifier: RO-C-MIDAS-5-PRL-TO-EXT3-V2.0, Server \newline\url{https://www.cosmos.esa.int/web/psa/ftp-browser\#rosetta}\label{ESA/PSA address_previous}}) and update it to V6 by adding more particles, correcting several manual errors, and adding newly developed shape descriptors (\citealp{Kim_Mannel_2022EGU}; Kim et al. in preparation).

This update is extremely important because all scientific data and statistics about the particle (e.g., its location, size, and height) were extracted from the masked areas, which were stored as separate data products in the archive. In the following, we present the most important entries in the particle catalog for this project.\newline

\noindent\textbf{Particle ID}:  a unique identifier for each particle that consists of the start time of the scan containing the particle (in the format year-month-day T hour minutes seconds), a particle number corresponding to the mask number in its mask file (e.g., each particle was given its own pixel mask so that one mask file could label many particles), and the target number on which the particle was detected. For example, the particle ID 2015-12-12T144849$\_$P01$\_$T11 in the catalog shows the scan on which the particle was detected started on the day 12 Dec. 2015 at the time 14:48:49. The number after "P" in the particle ID indicates the number of the mask, which is the number of the dust particle marking starting with one in each scan. The last entry in the particle ID is the number of the target on which the particle was detected (e.g., T11 for a particle on target 11).\newline

\begin{figure}
\includegraphics[width=\textwidth/2, height=4.3cm]{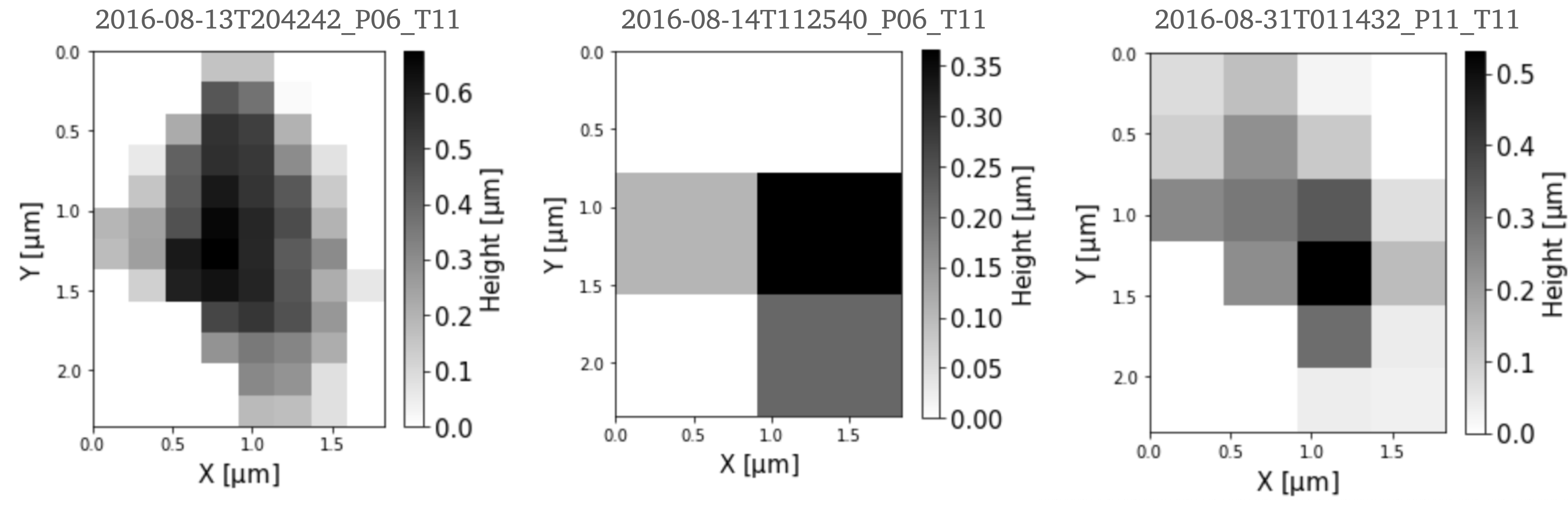}
\caption{Three scans of the same particle (particle ID of the left image: 2016-08-13T204242$\_$P06$\_$T11, of the middle image: 2016-08-14T112540$\_$P06$\_$T11, and of the right image: 2016-08-31T011432$\_$P11$\_$T11) with the master ID 2015-12-14T155236$\_$P02$\_$T11. For our studies, we selected the left image as the best representation because it has the best resolution (194.51~nm compared to the 781.84~nm and 389.01~nm resolution for the middle and right image) and shows the earlier scan, meaning that the risk of alteration during scanning is lower.}
\label{fig:ex_particle_selection}
\end{figure}

\noindent\textbf{Linked Particle ID}: describes a link between different observations of the same particle and provides the particle ID of the previous scan of the same particle. This is important because particles were often scanned several times (e.g., to optimize the scanning parameters). For further studies, such as the creation of dust coverage maps and clustering maps in Section~ \ref{sec:results_coverage_map} and the planned shape descriptor work, we only select the best representation of each particle. Figure ~\ref{fig:ex_particle_selection} describes an example of the selection for the best representation of each particle. \newline

\noindent\textbf{Master ID}:  gives the particle ID of the first time a particle was scanned. Thus, it is possible to obtain the particle IDs of all apparitions of the same particle via the same master ID.\newline

\noindent\textbf{Prescan Flag}:  indicates whether the presence of the visible and/or identified particle feature was already exposed in a previous scan. We note that only for $\sim$ 6\% (221/3523) of the particles' prescans are available. Since pre-scanning, dust collection, and search for dust particles on the target was heavily time-consuming and the mission time-limited, not all areas on the exposed targets could be pre-scanned (\citealp{bentley_lessons_2016}). However, all dust particles that were detected in scans with pre-scans were clearly distinct from the contamination pattern on the targets (flat, about 50-100 nm-sized bumps) and showed a typical morphology. Features larger than the typical contamination are expected to be cometary.
This flag also represents the special cases when there was no exposure of the target between the prescan and the scan in which the particle is seen, which likely have previously attached themselves to the cantilever tip during scanning of a particle and have subsequently fallen from the tip onto the target.\newline

\noindent\textbf{Particle Flag}:  indicates whether the particle is cometary in origin, based on several sets of criteria (e.g., not present prior to the previous exposure of the target, too big to be contamination, the clear "splatter" pattern of many small particles seen as fragmentation of larger particles on impact or during collection, etc.) or if a different origin is assumed (e.g., contamination on the target). In this study, we select and investigate only cometary particles. \newline

\noindent\textbf{Imaging Quality Flag}:  indicates whether a particle was imaged without issues. Examples of issues that caused a scan to be rejected include when the particle was not completely in the scanning range (e.g., a part of the particle lies outside of the boundaries of the image), parts of the particle are hidden by failed regions in the scan (e.g., a cantilever has reached maximum extension such that not the entire particle could be scanned), or strong artifacts appeared as, for instance, tip convolution\footnote[3]{As usual for AFM images, all MIDAS topographic images are a convolution of the tip and the particle surface shapes, with the strength of the effect depending on the condition of the tip (sharpness, shape, contamination) and the morphology of the particle. Note: a full-tip deconvolution was not possible due to the unexpectedly large cometary particles. For instance, MIDAS features a tip imaging target that allows the first 700~nm of the MIDAS tips to be imaged, however, the cometary particles imaged by MIDAS were mostly found to be much larger.}. We considered only those particles that were fully imaged without issues to determine their shapes in further studies.\newline

\noindent\textbf{Multiple Fragments Flag}: indicates whether the mask of the particles seems to contain multiple fragments that are lying over each other, such that a separation of the fragments was not possible. Such cases must be disregarded during further studies such as shape descriptor calculation to avoid falsification of the determined particle shape. Thus, we flagged multiple fragments when there were clearly multiple fragments contained within the particle mask or identification that were too hard to separate.\newline

\noindent\textbf{Trust Height Flag}:  indicates whether the measured height of the particle (i.e., defined as the difference between the highest pixel in the mask and the median of the heights of all pixels surrounding the particle mask) can be trusted. If the particle is not fully imaged or has one of a variety of imaging issues (e.g., artifacts, failed imaging covering part of the particle due to the reached maximum or minimum cantilever extension, a too-low image resolution), the estimation of the particle height may not be reliable. Furthermore, the majority of not fully imaged particles, possibly caused by the saturation of the maximal cantilever extension, were found to be larger than about 1.2 micrometers. Thus, here we define 1.2 micrometers as having "poor resolution," meaning that the height of the particle cannot be trusted. We select only particles with trusted height for further study because the height is expected to be a crucial input value for the size distribution, dust coverage, and clustering maps, as well as the calculation of shape descriptors.\newline

\noindent\textbf{Scientific data}: to date, these entries mostly indicate the area, the height in pixels, and micrometers, and the volume of the particle in cubic micrometers. The height information (e.g., height and the mean height of the surrounding area of the particle, the minimum and maximum height of pixels in the mask in micrometers) together with the meta data about the X-axis and Y-axis resolution (step size) of the scan, the number of steps, the X-axis and Y-axis coordinate of the first pixel in the scan (top left pixel in all images) in target coordinates, as well as the X and Y length of the image in micrometers, allow the 3D topography of the particles to be recreated. Together with the known location of each particle, this opens the opportunity to draw MIDAS dust collection maps solely using the particle catalog. \newline 

In addition to the scientific data, the catalog includes the tip number and target number used in scanning, the filename and start time of the last image of the tip made on the tip imaging target, as well as a bounding box containing the pixel coordinates which can be used to extract individual particles and masks from the image and mask files.\newline

\noindent The MIDAS particle catalog provides a unique and useful tool for identifying and investigating the structural properties of micrometer-sized cometary dust particles. Most importantly, the catalog allows quick access to a variety of particle characteristics, properties, and statistics. Furthermore, the catalog is a resource for future studies of cometary dust such as the development of MIDAS particle shape descriptors (Kim et al. in preparation). The most recent version of the catalog (discussed here) can be found in the ESA's Planetary Science Archive (PSA)\footnote[4]{{ESA planetary science archive of MIDAS/Rosetta (PSA 6.2.4), dataset identifier: RO-C-MIDAS-5-PRL-TO-EXT3-V3.0, \newline
\url{https://archives.esac.esa.int/psa/\#!Table\%20View/MIDAS=instrument}}\label{ESA/PSA address}}.

\subsubsection{Basic statistics of MIDAS particles}\label{sec:results_Basic_statistics}

The updated version of the MIDAS catalog V6 includes 3523 MIDAS dust particle identifications in total. The selection of particles usable for further investigation falls into two main steps: i) particle selection by using the particle catalog and additional analysis of the MIDAS AFM scans and ii) duplicate removal to ensure only the best representation of the particle among multiple scans of the same particle is used. Details can be found below.\\

\noindent For the particle selection, only particles of cometary origin and with a preferably low amount of alteration and thus clear appearance are of interest. Thus, we used the following selection criteria: Particles were probably cometary in origin (particle flag = 1) and imaged such that their height can be derived, for instance, exhibiting no strong image artifacts (trust height  flag = 1) or fully imaged particle (imaging quality  flag = 1). Furthermore, particles were not susceptible to being multiple fragments (multiple fragments flag = 0) and not suspected to have fallen off a tip, meaning that they were previously scanned, picked up with the tip, and dropped at different locations (prescan flag $\neq$ 4). Lastly, particles were not shown to be fragments of particles previously scanned (linked particle flag $\neq$ 3). 

With the particle selection work, we choose 2088 particles (493 particles before perihelion + 1595 particles after perihelion) that are usable for shape descriptor investigation in further analysis. A summary of how many particles fall within which selection criteria for each target together with their total number is given in Table~\ref{table:summary_how_many_particles}. \newline

\begin{table}[]
\caption{Number of MIDAS particles depending on selection criterion for each target.} 
\label{table:summary_how_many_particles}
\renewcommand{\arraystretch}{1.2}
\centering\begin{tabular}{lccccc}
\toprule
\multirow{2}{*}{\textbf{Selection criterion}}    & \multicolumn{5}{c}{\textbf{Target}}         \\ \cline{2-6} 
                                        & \textbf{all}          & \textbf{11}   & \textbf{13}    & \textbf{14}     & \textbf{15}      \\ \hline
Cometary origin                         & 3199 & 705 & 104 & 2377 & 13  \\
Not multiple fragment                   & 3170 & 568 & 137 & 2443 & 17  \\
Fully imaged                            & 2780 & 507 & 108 & 2148 & 15  \\
Height trusted                          & 2421 & 497 & 108 & 2443 & 10  \\
Not fallen off the tip, \\ no fragment of prior scan & 3347 & 759 & 116 & 2451 & 16       \\ \hline
All conditions                          & 2088 & 410 & 73   & 1595 & 10         \\ 
\bottomrule
\end{tabular}
\end{table}             

\noindent With regard to duplicate removal, we note that many MIDAS dust particles were scanned several times. The scientific reason for the duplication was to optimize the scanning parameters to obtain a better image with the highest possible resolution and optimized scanning area. However, we find that the dust particles often show signs of alteration from the scanning process (e.g.,~fragmentation due to a breaking-off of subunits and dragging them along the scanning direction). Thus, choosing the best image of a particle is a trade-off between optimal scanning parameters and a low amount of alteration as a result of scanning. Deciding on a case-by-case basis which scans of a particle are the best usable for shape descriptor development is a cornerstone of this study. In general, we chose the particle representation with higher resolution and/or the earlier scan (showing or at least implying the lowest amount of alteration or scanning artifacts). We find that in total 1857 particles (418 particles before perihelion~+~1439 particles after perihelion) are unique within 2088 particles. Thus, we used 1857 particles for further investigation. Based on the selection process described above, the detailed list of best representations of each particle depending on targets can be found in Table \ref{table:Best representation of 1857 particles} of Appendix~ \ref{appendix: Best representation of each particle}. \newline

\noindent In terms of the statistics of MIDAS particles, Table~\ref{table:summary_all_particles} shows how many  unique particles were collected in total depending on the exposure period for each target. It also contains two equivalent radii (i.e.,~defined as the radius of a disk or sphere with the same area or volume) and the volume of MIDAS particles together with some simple statistics. The standard dust equivalent radius on the 2D projection of the particles is calculated via Gwyddion and Pygwy\footnote[5]{Pygwy is a Python binding to Gwyddion's objects, methods, and functions, \url{http://gwyddion.net/documentation/user-guide-en/pygwy.html}} as it was done in previous publications, for instance,~\cite{bentley_morphology_2016}. However, we find that most MIDAS particles after impact show a flattened conical or cylindrical shape with an elongation~(\citealp{bentley_morphology_2016}), thus deviating from the simple 2D projection of a sphere assumed for the 2D equivalent radius. Therefore, we suggest improving the calculation of the equivalent radius of the particles before impact by taking into account available 3D information via the volume of MIDAS particles. For each particle, we first calculate the mean height of all surrounding pixels and subtract this value from all the height values in the mask to set the zero level and thus arrive at proper height values. Next, we multiplied the measured height of a pixel by the area of the pixel and sum over all pixels to measure the volume of all MIDAS particles. Finally, we calculated the 3D equivalent radius through the mean volume of the deposited particle on the assumption that the original particle was a sphere and was flattened by impact without any mass loss. In conclusion, the 3D equivalent radius presumably gives a better impression of the initial size of the particle before it hit the target than the 2D equivalent radius based on the 2D projected area of the particles. We note that the volume mean could systematically result in a smaller estimate of the radius since it does not account for the lowered particle volume into the surface of the target, in contrast to the estimate of the lower limit radius based on a 2D projection.

\begin{table*}[]
\caption{Number of particles collected depending on the period for each target.}
\label{table:summary_all_particles}
\renewcommand{\arraystretch}{1.2}
\centering\begin{tabular}{llllll}
\toprule
\textbf{Target}  & \textbf{Target 11}    & \textbf{Target 13}    & \textbf{Target 14}   & \textbf{Target 15}   & \textbf{All}    \\\hline
\textbf{Period}   & \begin{tabular}[c]{@{}l@{}}Sep. 2014 - \\ Nov. 2014\end{tabular} & \begin{tabular}[c]{@{}l@{}}Dec. 2014 - \\ Feb. 2015\end{tabular} & Feb. 2016   & \begin{tabular}[c]{@{}l@{}}Feb. 2015 - \\ Mar. 2015\end{tabular} & \begin{tabular}[c]{@{}l@{}}Sep. 2014 -\\ Feb. 2016\end{tabular} \\\hline
& Pre-perihelion    & Pre-perihelion    & Post-perihelion    & Pre-perihelion    &    \\\hline
\begin{tabular}[c]{@{}l@{}}Number of particles scanned \\ usable for shape descriptor\end{tabular}    & 410    & 73    & 1595    & 10    & 2088    \\\hline
\begin{tabular}[c]{@{}l@{}}Number of unique particles \\ usable for shape descriptor\end{tabular}    & 350    & 61    & 1439    & 7    & 1857    \\\hline
\begin{tabular}[c]{@{}l@{}}Particle equivalent radius \\on the 2D projection\\mean $\pm$ standard deviation [$\mu$m] \end{tabular}    & 0.92 $\pm$ 0.75    & 0.64 $\pm$ 0.64    & 0.91 $\pm$ 0.81    & 1.61 $\pm$ 0.93    & 0.91 $\pm$ 0.79    \\\hline
\begin{tabular}[c]{@{}l@{}}Particle equivalent radius \\ derived from volume \\mean $\pm$ standard deviation [$\mu$m] \end{tabular}    & 0.55 $\pm$ 0.42   & 0.37 $\pm$ 0.35    & 0.57 $\pm$ 0.46   & 0.95 $\pm$ 0.50    & 0.56 $\pm$ 0.45    \\\hline
\begin{tabular}[c]{@{}l@{}}Volume of particle\\mean $\pm$ standard deviation [${\mu}m^3$] \end{tabular}    & 2.44 $\pm$ 7.00 & 1.11 $\pm$ 3.48 & 3.40 $\pm$ 14.89 & 6.21 $\pm$ 6.77 & 3.15 $\pm$ 13.49\\
\bottomrule                                                        
\end{tabular}
\end{table*}

We find that the calculated equivalent radii range from about 80~nm to 8.5~$\mu$m. In particular, the mean equivalent radius derived from the 2D projection of the particles measures 0.91~$\pm$~0.76~$\mu$m, which is larger compared to the value derived by the volume 0.56 $\pm$ 0.45, showing overlapping results in the range of the uncertainty. This may indicate either MIDAS particles are flattened by impact already, thus showing a larger particle size in the 2D projection or particles have a natural elongation and preferentially stick with their longer side on the targets (\citealp{bentley_morphology_2016}).

Comparing the equivalent radii of particles collected pre-perihelion (mostly collected on target 11, few particles on targets 13 and 15) to those radii derived for particles from post-perihelion (target 14, all particles originating from the outburst on February 16, 2016; \citealp{Gruen_2016_outburst}) shows overlapping results in the range of the uncertainty as well. We do not find that the calculated values themselves show a noticeable difference: both the mean values of the 2D projection-based and volume-based equivalent radius are similar for pre-perihelion and the outburst. Taking those facts together, particles have similar projected areas and volume regardless of cometary activities, leading to the conclusion that different cometary activities (e.g., nominal activity or outburst) are not related to the either size of the particle or the degree of compression.

The mean deposition velocities measured by the GIADA instrument for the related MIDAS collection periods are lower for pre-perihelion (about 3~m~s$^{\rm -1}$ for pre-perihelion and 7 m s$^{\rm -1}$ for the outburst; \citealp{Longobardo_2020a_Merging_data}). Thus, we may conclude that the difference in impact velocity between 3~m~s$^{\rm -1}$ and 7~m~s$^{\rm -1}$ is not the main driver for different particle sizes in MIDAS particle collection. Furthermore, dust source difference ~(\citealp{Longobardo_2020a_Merging_data}) seems to be unrelated to the dust particle size as well because particles collected in targets 11 and 14 are expected to stem from different cometary terrain types. Although \citet{Longobardo_2022_MNRAS} observed a dust density variation (e.g., higher density particles $\rho$ > 1000 g\,cm$^{\rm -3}$ during the time MIDAS collected the particles on target 11 pre-perihelion, but lower density particles $\rho$ < 1000 g\,cm$^{\rm -3}$ during the time MIDAS collected the particles of the outburst on target 14) that can be explained by different dynamics of lighter and heavier particles (slower velocities for heavier particles, thus more spread in the coma, hence lower detection rates), MIDAS particle sizes seem not to be affected by density differences. Thus, we conclude that different cometary parameters (e.g., cometary activities with different impact velocities) do not lead to a different behavior of fragments of parent particles during deposition, which shows consistency with previous studies (e.g., \citealp{Longobardo_2022_MNRAS}).

According to \citet{Fulle_2020_ice_sublimation}, micrometer-sized particles are too small to be ejected from a cometary source by simple gas pressure. This finding is in agreement with the observation of the COSIMA instrument, detecting a lack of tens of micrometer-sized or smaller dust particles in the coma of comet~67P (\citealp{merouane_dust_2017}). Thus, it seems probable that the size of the particles observed by MIDAS is mainly created by fragmentation, which should depend on the initial size before impact, impact speed (\citealp{Ellerbroek_labstudy_2017}), and the interconnected properties such as density and porosity, as well as cohesion~(\citealp{Lasue_simulation_2019}). Thus, it should be noted that all size distributions discussed in Sect.~\ref{sec:results_size_distribution} may not reflect the particle sizes of cometary dust in the coma but, rather, the particle sizes after collection by MIDAS. 

Lastly, we compare the measured volume of MIDAS particles to the calculated volumes of commonly found shapes. Table~\ref{table:various shapes of volume and radius} shows the measured volume, along with the volumes calculated for a cone and a cylinder with the same radius as the mean equivalent radius derived from the 2D particle projection, as well as the derived equivalent radius based on the volume. We find that the mean value of the volume of MIDAS particles (3.15 $\pm$ 13.49~$\mu$m$^{\,\rm{3}}$) lies between the mean value of a cylinder shape (8.81~$\pm$~42.69~$\mu$m$^{\,\rm{3}}$) and a cone shape (2.94~$\pm$~14.23~$\mu$m$^{\,\rm{3}}$), but is more similar to a cone shape. The cone shape slightly underestimates the volume, thus the average real shape might be close to a cone shape but with a rounded tip. Since we did not find any other rather irregular shape of MIDAS particles, we conclude that the natural appearance of MIDAS particles resembles a pyramid and more like an upper segment of an elongated ellipsoid (elongated by 2.87$^{+1.90}_{-0.44}$, lying flat on the target, as found by~\citealp{bentley_morphology_2016}). 


\begin{table*}[]
\caption{Measured volume, the calculated volumes on the assumption of cone-shaped and cylinder-shaped MIDAS particles, and the resulting 3D equivalent radius.} 
\label{table:various shapes of volume and radius}
\renewcommand{\arraystretch}{1.2}
\centering\begin{tabular}{llll}
\toprule 
\textbf{}  & \textbf{Measured}    & \textbf{Cone shape}    & \textbf{Cylinder shape}\\\hline
Volume\\mean $\pm$ standard deviation [$\mu$m$^{\,\rm{3}}$]  & 3.15 $\pm$ 13.49    &  2.94 $\pm$ 14.23   & 8.81 $\pm$ 42.69\\\hline

Radius derived from volume\\mean $\pm$ standard deviation [$\mu$m]  & 0.56 $\pm$ 0.45    &  0.51 $\pm$ 0.45   & 0.73 $\pm$ 0.65\\
\bottomrule
\hline                                                         
\end{tabular}
\end{table*}

\subsubsection{Dust coverage maps}\label{sec:results_coverage_map}

Based on the selection of the best representation of each dust particle, that is, the removal of duplicates within the updated particle catalog, as described in Section~\ref{sec:results_Basic_statistics}, we create dust coverage maps. These are extremely useful visual aids to assess not only the spread of dust particles on the MIDAS targets but also to check the results of the clustering algorithm. The maps can also be used to derive results, such as the percentage of the scanned area, the area covered by particles, ratios of numbers of fragments versus parent particles, and so on (see Section~\ref{sec:results_Fragmentation}). 

We prepare dust coverage maps of the four different MIDAS targets with cometary dust detection (i.e., targets 11, 13, 14, and 15) for all particles (without duplicates) in 2D and 3D. For each dust coverage map, we provide two versions, namely: with and without particles that are not fully imaged (imaging quality = 1 for entirely imaged particles, 2 for particles not imaged completely). In particular, dust coverage maps with all selected particles, regardless of whether they are fully scanned or not, were created because parent particles are expected to be too large to be fully scanned; thus, a relation between fragments and their parent may only become visible in a map including partially scanned dust particles. We also created maps with an indication of the scanned area, where for overlapping scans only the bounding box of the scanned area is shown to avoid overcrowding of the plot. This is highly useful because MIDAS could only scan its collection targets in stripes that were separated by about 80~$\mu$m-sized inaccessible areas vertically and horizontally, stitched together by patches of maximally 80~$\mu$m x 80~$\mu$m-sized scans~(\citealp{bentley_lessons_2016}). Thus, a seeming cluster must always be checked against the scanned area around it to avoid false cluster identification. Additionally, we investigated 3D versions of the maps to allow a better assessment of the particle heights. The maps are in general 2D grayscale images showing the exact particle topographies at their respective locations on the target, where darker color means higher height. 

Examples of the dust coverage maps (e.g., 2D grayscale images with the indicated scanned area and 3D grayscale images) are presented in Appendix \ref{appendix_dust_coverage_maps}. The data used in this paper (e.g., mask data and the location of particles in the MIDAS particle catalog, scan files, and their location in the target) are available in the ESA/PSA\textsuperscript{\ref{ESA/PSA address}}.

\subsubsection{Size distribution of the MIDAS dust particles}\label{sec:results_size_distribution}

To investigate the MIDAS particle size distribution, we used 1857 unique MIDAS particles that are usable for further analysis. For each size distribution, we evaluate a series of shape descriptors and calculate both the equivalent radii of dust particles by their 2D projection and the one derived from the measured volume of MIDAS particles; we note that the latter  is expected to give a more accurate size of the dust particles (see Section~\ref{sec:results_Basic_statistics}). In the fit of the size distribution, we excluded the smaller bins (up to equivalent radii~$\leq$~500~nm) due to the resolution limit of the MIDAS instrument. This is because almost half of the individual scans have resolution limits above $\sim$ 500~nm, which causes particles smaller than 500~nm to be hardly distinguishable from noise. Thus, their number is possibly underestimated. Additionally, following the idea of hierarchical particles suggested by \citet{mannel_fractal_2016} and \citet{blum_evidence_2017}, setting free such small particles should be a rare event, for instance, that is only to occur at high-impact velocities or especially loosely bonded small subunits. In conclusion, a low number of nanometer-sized individual particles is expected due to scarcity and detection bias.

\begin{figure}
\includegraphics[width=9cm, height=6.5cm]{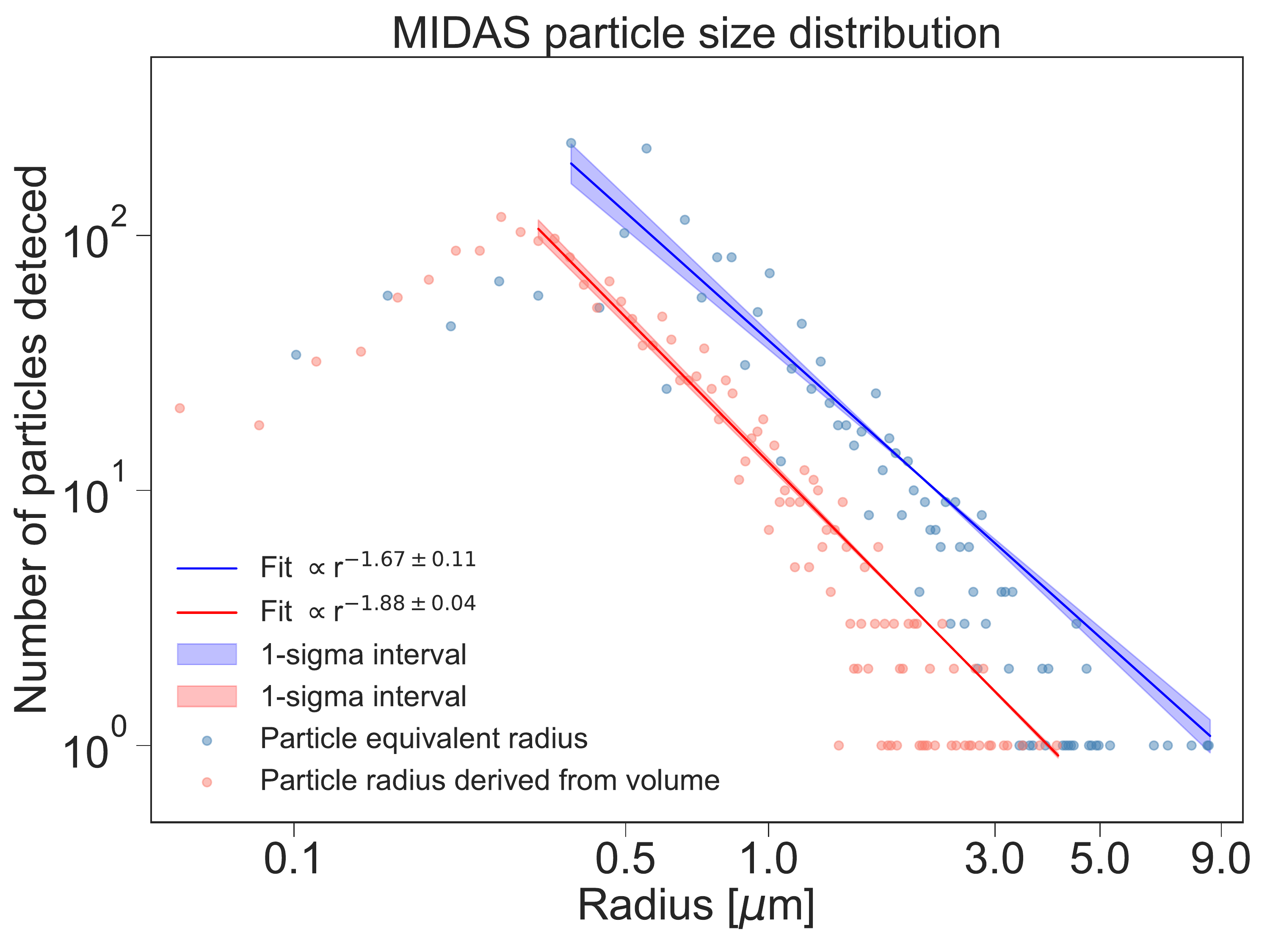}
\caption{MIDAS particle size distributions with equivalent radii based on the particle's 2D projection (blue line) combined with a power law ${a \cdot r}^{\rm{\,b}}$ fit with an index $\rm{\,b}$ of $\sim$ -1.67 and the one derived from volume (red line) with an index $\rm{\,b}$ of $\sim$ -1.88.}
\label{fig:grainsize_distribution}
\end{figure}

Figure \ref{fig:grainsize_distribution} shows the size distribution based on the equivalent radii of the 2D projection (blue line) and the one derived from the volume (red line) of the MIDAS particles, the error stated is one sigma. We derive that the size distribution of the MIDAS particles can be described with a power law, ${r}^{\rm{\,b}}$, fitted with an index $\rm{\,b}$ of -1.88 $\pm$ 0.04 (volume based) to -1.67 $\pm$ 0.11 (projection-based). In particular, we find that the data are similar for radii larger than about 1~micrometer, but the slope of the distribution based on the 2D projection of the particles is slightly steeper and contains a larger scatter in the data points, whilst the radii based on the volume show less spread and the fitted slope is shallower. Furthermore, we find a slight underpopulation of larger-size bins, where the low statistics leads to a larger uncertainty of the fitted slope. An explanation could be a fragmentation mechanism during collection, where larger (weaker) particles fragment more frequently than smaller ones.

\begin{center}
\begin{figure*}
\centerline\centering\includegraphics[width=18.5cm, height= 7cm]{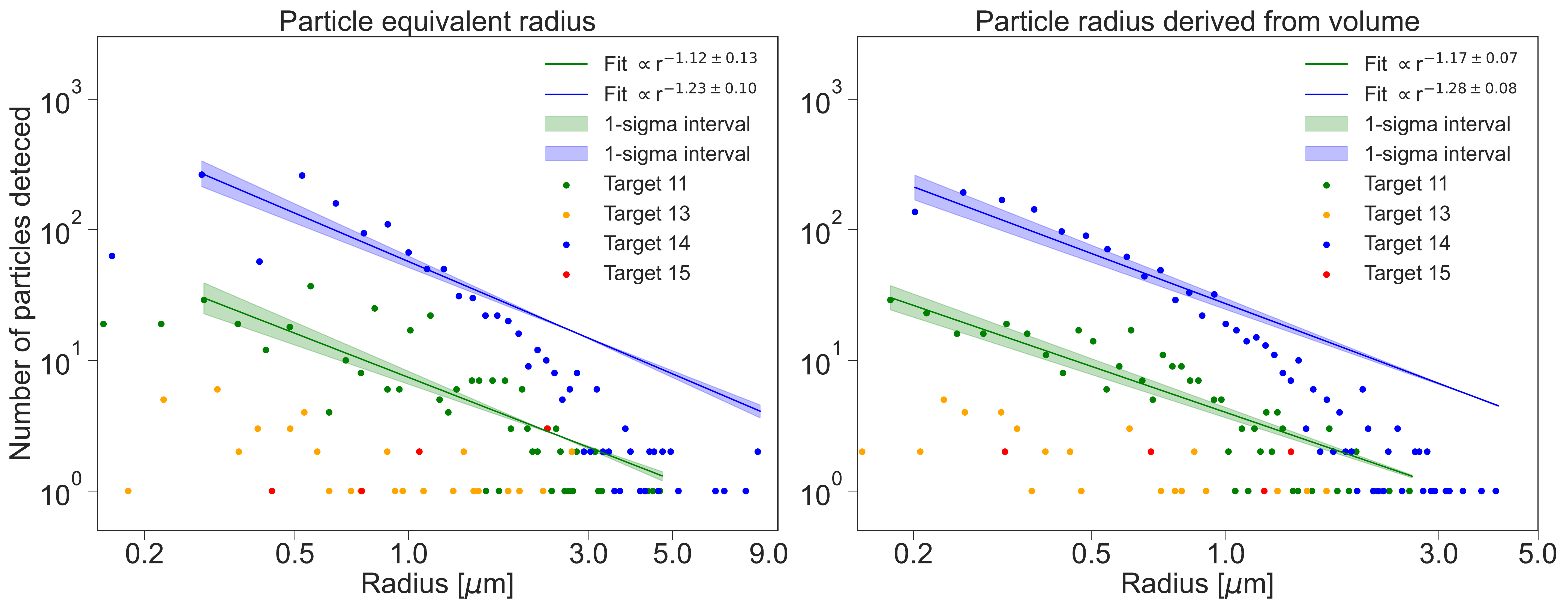}
\caption{\textbf{}Size distributions of the equivalent radius based on the 2D projection depending on targets combined with a power law ${a \cdot r}^{\rm{\,b}}$ fit (green and blue lines for target 11 and 14) with an index $\rm{\,b}$ of $\sim$ -1.12 to -1.23 (left). \textbf{}Size distributions of the equivalent radius derived from volume depending on targets combined with a power law ${a \cdot r}^{\rm{\,b}}$ fit (green and blue lines for target 11 and 14) with an index $\rm{\,b}$ of $\sim$ -1.17 to -1.28 (right).}
\label{fig:grainsize_distribution_targets}
\end{figure*}
\end{center}

\begin{figure*}
\centerline\centering\includegraphics[width=18.5cm, height= 7cm]{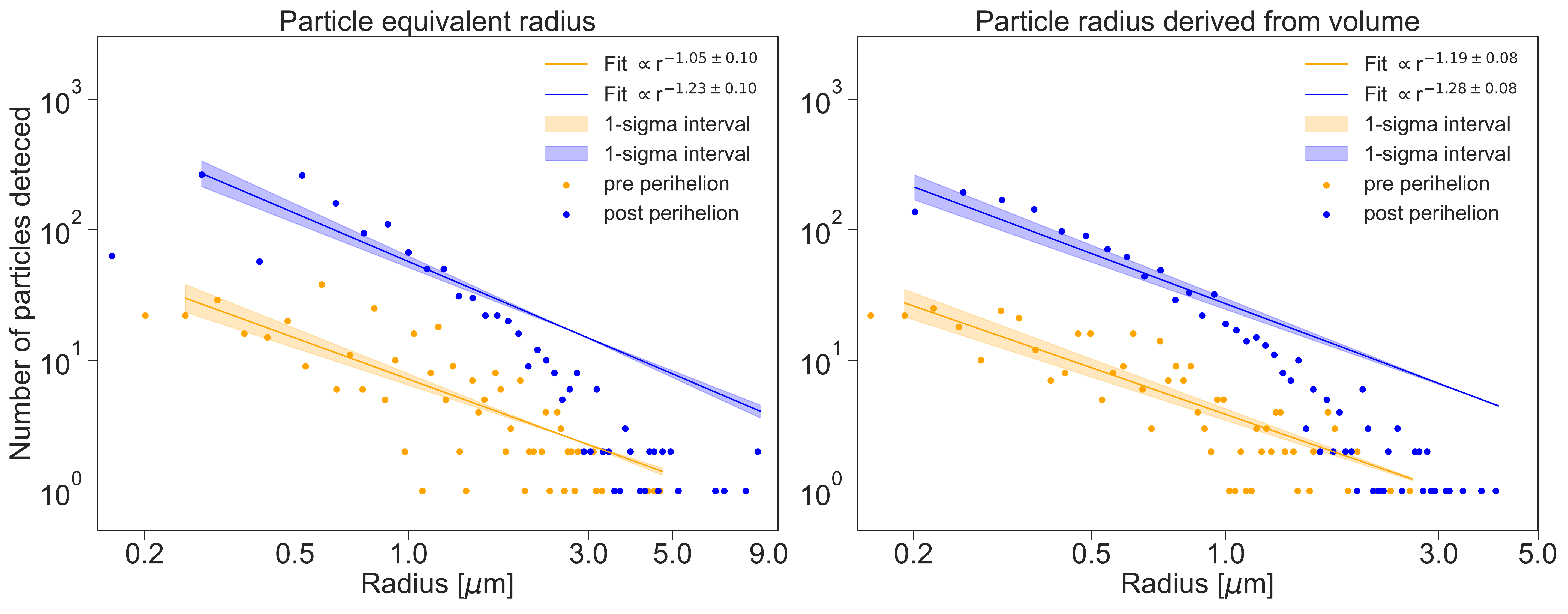}
\caption{\textbf{} Size distributions of the equivalent radius based on the 2D projection depending on collection periods combined with a power law ${a \cdot r}^{\rm{\,b}}$ fit (orange and blue lines for pre- and post-perihelion) with an index $\rm{\,b}$ of $\sim$ -1.05 to -1.23 (left).  Size distributions of the equivalent radius derived from volume depending on collection periods combined with a power law ${a \cdot r}^{\rm{\,b}}$ fit (orange and blue lines for pre- and post-perihelion) with an index $\rm{\,b}$ of $\sim$ -1.19 to -1.28 (right).}
\label{fig:grainsize_distribution_periods}
\end{figure*}

Figures~\ref{fig:grainsize_distribution_targets} and \ref{fig:grainsize_distribution_periods} show the size distributions with the equivalent radius on the 2D projection of MIDAS particles (left) and the one derived from the volume (right) depending on targets and different periods, respectively, and the error stated is one sigma. We find that the slope of the size distributions with the equivalent radius based on the particle's 2D projection shows a slight difference for different targets and periods. To be exact, the post-perihelion data set is coincident with the target 14 dataset, thus there is no difference in the fitting result. The pre-perihelion data set is consisting of the data from targets 11, 13, and 15, however, targets 13 and 15 have low statistics, thus the fitting result of target 11 is not much different from that for pre-perihelion (a power law index of -1.12 $\pm$ 0.13 for target 11 and of -1.05 $\pm$ 0.10 for pre-perihelion). Investigating the size distribution of equivalent radius based on the 2D projection and the one derived from the particle's volume, we find overlapping uncertainties for all cases for pre- and post-perihelion. The relative similarity of the slopes of the size distributions for each collection period may imply that different cometary activities and impact velocity may not play the driving role in the fragmentation of larger particles, but only in the number of collected parent particles. This confirms the findings from Table~ \ref{table:summary_all_particles} (see also Sect. \ref{sec:results_Basic_statistics}). 

We note that the equivalent radii based on 2D projection follow the fits less well than those based on the volume. In particular, there is a slight overpopulation of smaller particles and an underpopulation of radii larger than a micrometer. We suggest that the deviation for the 2D projection case is correlated to different compaction processes at different size scales, which again implies inevitable dust alteration by impact.

\subsection{Fragmentation and MIDAS parent particles (cluster)}\label{sec:results_Fragmentation}

While the previous sections discussed the properties of the dust particles and fragments in the particle catalog, in this section, we described how we use the data to discuss potential parent particle sizes and fragmentation mechanisms. 

\subsubsection{Dust collection process}\label{sec:Dust_collection_process} 
The main question in the present section is whether it is possible to deduce the arriving MIDAS parent particle size before fragmentation. The idea is to roughly understand dust particle deposition on the MIDAS targets and relate it to the cluster size found by the mean shift clustering algorithm described in Section~\ref{sec:clustering algorithm}. 

The size of the clusters determined by the algorithm strongly depends on the $h$ value (the bandwidth) in Equation~\ref{mean_clustering}. As the choice of $h$ is crucial, but since the expected MIDAS cluster size is unknown, we orientate on the results of other Rosetta dust instruments and laboratory experiments to set a reasonable value. The bandwidth should be a strong function of the impact velocities as well as initial particle sizes and material strengths (\citealp{Ellerbroek_labstudy_2017, Lasue_simulation_2019}). However, \citet{Ellerbroek_labstudy_2017} found that a large mass loss of the parent particle on impact can be expected in certain cases. In particular, they developed a mass transfer function TF, which is defined as the ratio between the mass of the post-impact deposit ($M_{\rm{\,post}}$) and the mass of the particle pre-impact ($M_{\rm{\,pre}}$), that is: TF~=~$\frac{M_{\rm{\,post}}}{M_{\rm{\,pre}}}$. 

For larger particles (80~$\mu$m < $d_{\rm{\,pre}}$ < 410~$\mu$m), which are comparable to the dust particle size investigated by COSIMA and GIADA, they observed two different cases depending on the impact velocity. At low velocities ($v$ < 2 m s$^{\rm{-1}}$), the majority of these particles leave deposits with a "shallow footprint" morphology, namely, a circular or ellipsoidal line of small fragments encompassing a sparsely covered or even empty area with a typical TF < 0.2 (i.e., less than 20 percent of the mass of the parent particle was observed to be deposited on the target). Around 10$\%$ of the incoming particles leave a "single" morphology, where most of the parent particle mass is expected to have stuck on the target and the particle seems to be without (major) alteration. At velocities above 2 m s$^{\rm-1}$, a "pyramid-shaped" deposit is left on the target with 0.2 < TF < 0.5, that is, maximally 50 percent of the mass of the incoming particles was deposited on the targets. The deposit looks as if the particle structure partially broke down, scattering fragments around the bulk of the particle.

Particles smaller than 80~$\mu$m seemed to stick without alteration (creating "single" morphologies). It is possible that dust particles that are just small enough can stick without (strong) alteration, in particular without large mass loss, however, it is not known from which sizes this effect takes place for real cometary particles because the material composition of cometary dust is expected to be different from that used in~\citet{Ellerbroek_labstudy_2017}, that is, pure SiO$_{2}$. Furthermore, the 80~$\mu$m threshold might be biased by the resolution limit for the determination of the size and, hence, the estimated mass of the incoming particles. Nevertheless, the observations of ~\citet{Ellerbroek_labstudy_2017} suggest that if the size of MIDAS parent particles is larger than a certain value, potentially around 80~$\mu$m but maybe also much smaller, a substantial loss of parent particle mass, that is, a TF value much smaller than 1, has to be taken into consideration. 

A further reduction of the actually detected particle mass by MIDAS may be caused by technical limitations. MIDAS could only fully image particles with heights much less than 10~$\mu$m and it could only scan its targets in about 80~$\mu$m high stripes separated by about 80~$\mu$m high areas~(\citealp{bentley_lessons_2016}). If fragments of a parent particle are higher than 10~$\mu$m or were scattered in an unaccessible target area, they were not detected. Furthermore, particle loss of particles already stuck to the MIDAS target seemed to occasionally happen at the Flight Spare unit (a structurally identical MIDAS instrument operated on Earth), especially when changing the target located in front of the AFM unit.

\subsubsection{Retrieval of the MIDAS parent particle sizes}\label{sec:Retrieval_of_the_MIDAS_parent_particle sizes} 

We carried out two complementary approaches to the approximate idea of the MIDAS parent particle size. In both cases, we used the clusters determined by the algorithm described in Section~\ref{sec:clustering algorithm}. The first one is based on a visual comparison of the patterns in the MIDAS clusters, in particular, when they are similar to the shallow footprint dust deposit structures found in the laboratory experiment of ~\citet{Ellerbroek_labstudy_2017}. The second one attempts to calculate the parent particle size from the total volume deposited in the clusters.  

\begin{figure}
\centerline\centering\includegraphics[width=9cm, height=9cm]{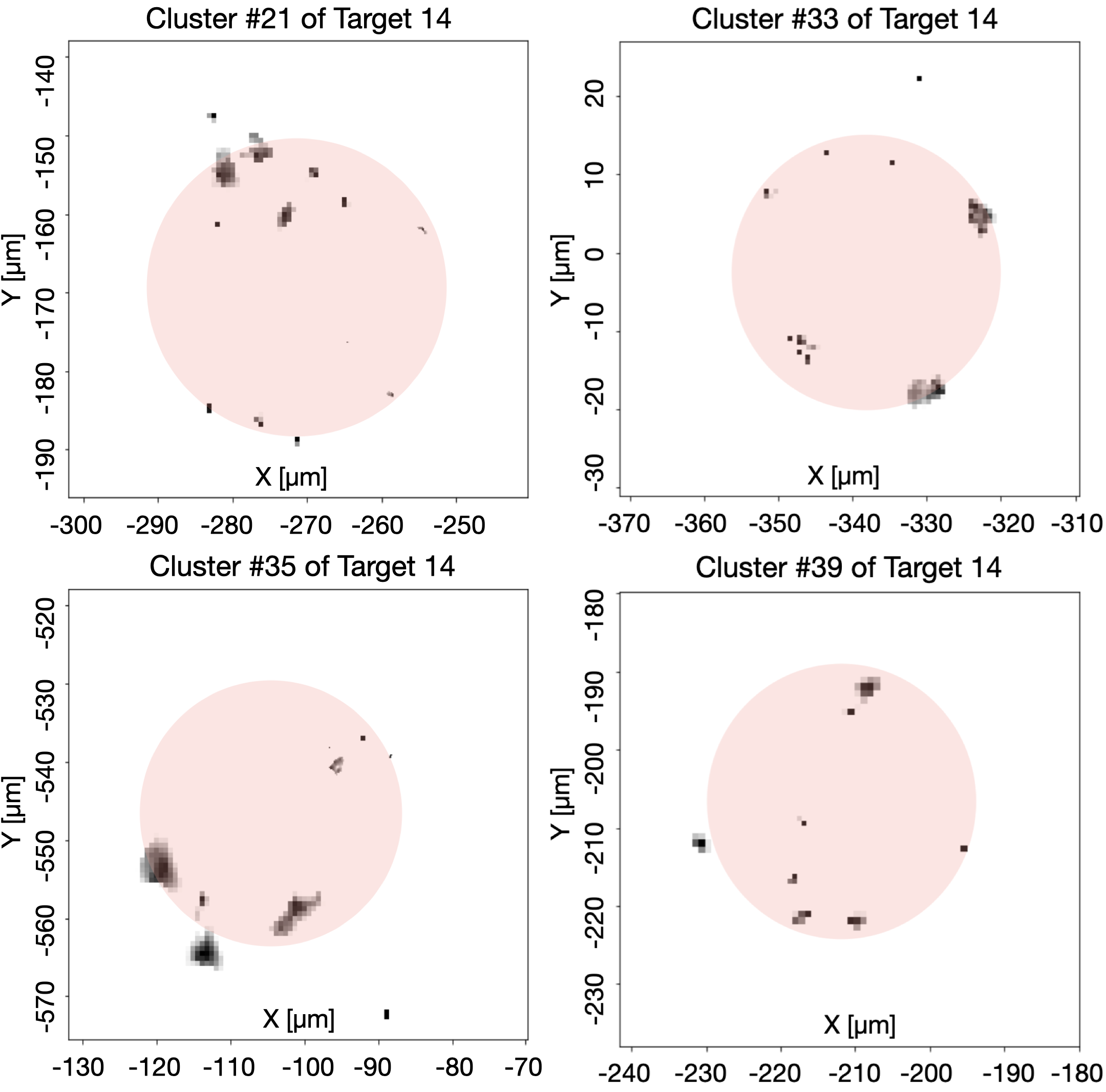}
\caption{Exemplary clusters of fragments deposited on MIDAS targets that show a shallow footprint-like morphology. The empty area encircled by the fragments is approximately marked with red circles and has a radius of about 15 - 25~$\mu$m.}
\label{fig:approximated_cluster_size}
\end{figure}

The shallow footprint morphology determined in the \citet{Ellerbroek_labstudy_2017} laboratory experiments stands out relatively well on dust collection targets since it shows a circular or ellipsoidal line around a (nearly) empty area. It is expected that a dust agglomerate arrived slowly and bounced on the target, leaving only some dust fragments falling off its rim. This implies that the size of the empty area should be roughly comparable to the parent particle size. A visual inspection of MIDAS dust collection targets shows features that look very similar to shallow footprints, for instance, as shown in Figure ~\ref{fig:approximated_cluster_size}. The empty area encircled by the fragments has a radius of about 15 - 25 ~$\mu$m micrometers. We note that these radii are just drawn by eye and not a result of the clustering algorithm. Judging from the shallow footprint-like MIDAS deposits, the parent particle size would fall more into the tens of micrometer regime with a mean size around 20~$\mu$m in radius. This would be well in agreement with the smaller end of COSIMA and GIADA measurements, for example, \citet{Merouane_COSIMA_size_D_2016, fulle_evolution_2016, ACLR2018, Longobardo_2020a_Merging_data}. We note that larger particles of a millimeter to a centimeter size are not part of the MIDAS data set due to the size of targets (see Sect. \ref{MIDAS_AFM}).

In a second, complementary and straightforward attempt, we calculated the MIDAS parent particle radius based on the assumption that a spherical parent particle entered MIDAS and deposited its whole mass as fragments on the target, similarly to the approach taken in~\citet{Merouane_COSIMA_size_D_2016}. In particular, summing up the volume of all fragments in one cluster, namely, adding the height (i.e., peak height above the surrounding background pixels) times pixel area for each cluster member, leads to the parent particle radius given in Table~\ref{table:retrieval of the MIDAS parent particle sizes}.

\begin{table}[]
\caption{Parent particle radius derived from the volumes of all particles in each cluster.} 
\label{table:retrieval of the MIDAS parent particle sizes}
\renewcommand{\arraystretch}{1.2}
\centering \begin{tabular}{lcc}
\toprule 
Target & \begin{tabular}[c]{@{}c@{}}Derived \\ parent \\ particle radius {[$\mu$m]}\end{tabular} & \begin{tabular}[c]{@{}c@{}}Mean and std of\\ parent \\ particle radius {[$\mu$m]}\end{tabular}\\
\hline 
11     & 0.32 – 6.28     & 4.68 $\pm$ 1.31     \\
13     & 0.45 – 4.11     & 3.07 $\pm$ 0.84     \\
14     & 0.58 – 9.86     & 7.34 $\pm$ 2.05     \\
15     & 1.01 – 3.40     & 2.59 $\pm$ 0.61  \\  
\bottomrule 
\end{tabular}
\end{table}

We find that the MIDAS parent particle radii derived from their volume as given in Table~\ref{table:retrieval of the MIDAS parent particle sizes} are 2.59 to 7.34 micrometers in size, which is smaller than the estimated few tens of micrometer parent particle radius derived from the diameter of shallow footprint-like MIDAS deposits. These small sizes of parent particles are not expected to be able to create the observed dust deposit patterns, especially not the shallow footprints. Thus, this finding confirms the observation in~\citet{Ellerbroek_labstudy_2017} that a vast majority of the parent particle mass is not transferred to the targets. It also emphasizes that MIDAS lost an unknown part of the mass of cometary parent particles during dust collection. In conclusion, the true MIDAS parent particle sizes can only be roughly estimated and the measured volume of MIDAS particles presents a lower limit of it.

\begin{figure*}
\centerline\centering\includegraphics[width=18.5cm, height= 7cm]{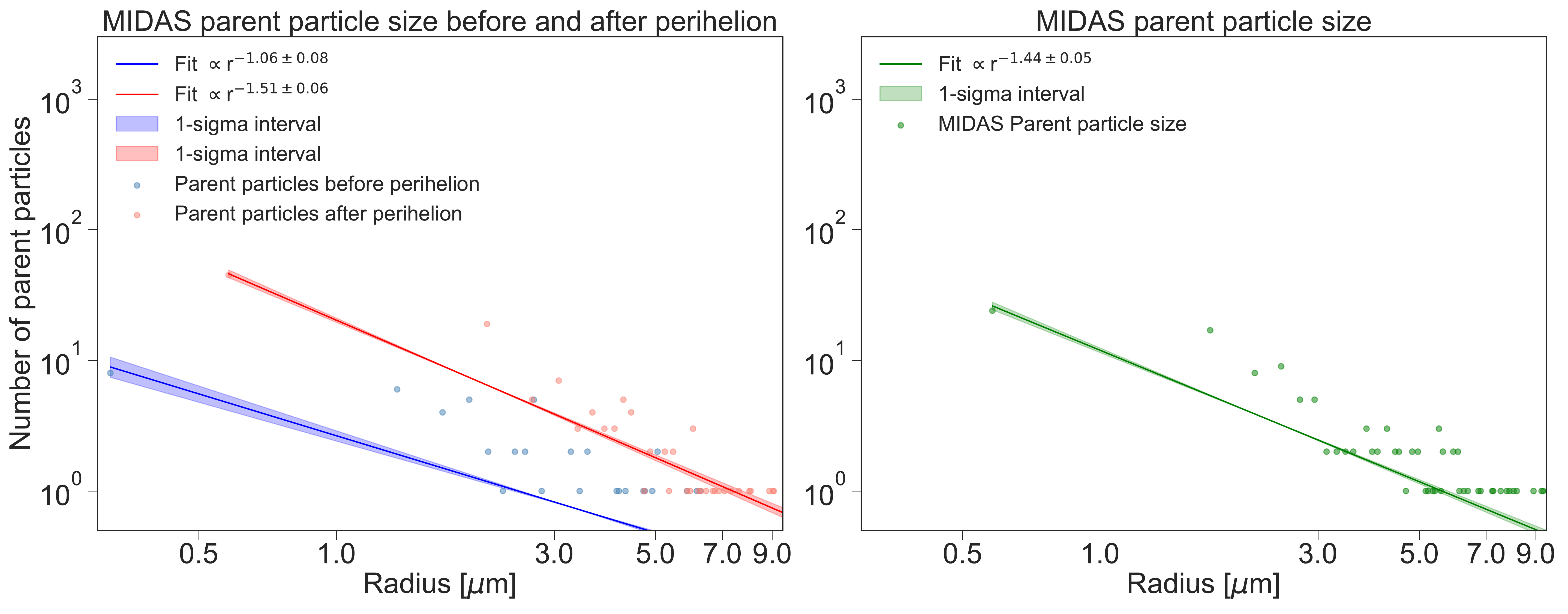}
\caption{MIDAS parent particle size distributions with a power law ${a \cdot r}^{\rm{\,b}}$ fit with an index $\rm{\,b}$ of $\sim$ -1.06 before perihelion (blue line), the one with an index $\rm{\,b}$ of $\sim$ -1.51 after perihelion (red line), and the one with an index $\rm{\,b}$ of $\sim$ -1.44 for whole MIDAS particles (green line). We note that the parent particle sizes are roughly estimated at lower limits and the uncertainty given in the graphs is only that of the fitting routine, hugely underestimating the real uncertainty of the size measurement.}
\label{fig:parent particle size distribution}
\end{figure*}

Figure \ref{fig:parent particle size distribution} shows the roughly estimated parent particles as size distribution before and after perihelion and whole MIDAS particles, the error stated is one sigma. We found the slope of MIDAS parent particle sizes before and after perihelion is similar (i.e., -1.06 to -1.51 with an unknown but large uncertainty attached to both slopes), which confirms the previous finding from Sect. \ref{sec:results_size_distribution} that the dust particles are similar for different collection periods. Furthermore, the slope of whole MIDAS parent particles shows $\sim$ -1.44, which indicates reasonable consistency with the finding from COSIMA for the smaller particles (e.g., the size distribution of smaller particles between 15 and 75 $\mu$m, which are comparable to MIDAS parent particle size ranges, has a power index of $\sim$ -1.6 to 2.8; \citealp{Merouane_COSIMA_size_D_2016, merouane_dust_2017}).


\subsubsection{Size of MIDAS clusters}\label{sec:The_size_of_MIDAS_clusters} 

To obtain sensible results about the MIDAS cluster size from the clustering algorithm a reasonable value for the bandwidth,~$h,$ in Equation~\ref{mean_clustering} must be chosen. In the previous section, we developed the idea that the MIDAS parent particles may have a roughly estimated size of a few tens of micrometers; thus, they seem to be similar in size to the smaller particles used in the laboratory study of~\citet{Ellerbroek_labstudy_2017}, and similar to the smaller particles detected by COSIMA and GIADA. Then, we decided to determine the $h$ value and the cluster size by an approximated cluster size of deposits found in~\citet{Ellerbroek_labstudy_2017}. However, \citet{Ellerbroek_labstudy_2017} showed that the degree of dust alteration and the cluster size by “hit and stick” collection depends not only on the parent particle size but also on impact velocity. As a result, the expected cluster size on the different MIDAS targets is assumed to be dependent on the estimated mean velocities of the particles hitting each target. Those mean velocities can be assumed from speed measurements of GIADA (\citealp{Longobardo_2020a_Merging_data}) taken for compact particles in the related MIDAS collection periods. 

Table~\ref{table:band_width_value_cluster} shows that targets 11 and 13, holding particles from early pre-perihelion (around lander delivery in November 2014), collected particles with relatively slow velocities around 3~m~s$^{\rm -1}$. Target 14 collected particles during the  February 16, 2016 post-perihelion outburst with slightly higher velocities around 7~m~s$^{\rm -1}$ (which is still a slow velocity compared to particle speeds around perihelion). Target 15 only holds a negligible amount of particles collected late pre-perihelion (March 2015) with an impact speed of roughly 7~m s$^{\rm -1}$. Those impact velocities are both in the range used by~\citet{Ellerbroek_labstudy_2017} and, thus, a visual inspection of the deposits in their study leads to expected cluster radii between 25~micrometers for slower deposited particles and 35~micrometers for faster-deposited ones. This seems to be a reasonable approach because the resulting variation is only a function of the relative velocity between the spacecraft and dust particles with the fixed MIDAS targets relative to the spacecraft movement and orientation (see Section \ref{MIDAS_AFM}), which might be possibly causing dust particle distribution shown in Figures \ref{fig:2D_dust_clustering_map_10} and \ref{fig:2D_dust_clustering_map_14} (i.e., not uniform across the shown area).

Table~\ref{table:band_width_value_cluster} gives an overview of the measured mean velocities of GIADA for the related collection periods of each MIDAS target with cometary dust, the determined cluster diameters based on the work of \citet{Ellerbroek_labstudy_2017}, and the applied bandwidth value,~$h$. For the chosen bandwidth value, $h$, we also carried out internal (e.g., k-nearest neighbors algorithm and silhouette analysis; \citealp{Fix1989, Rousseeuw1987}) as well as  external validations (e.g., comparison of the resultant cluster number based on the mean shift method to the one with the random distribution), which can be found in Appendix \ref{appendix: band_width}. Running the clustering algorithm described in Section~\ref{sec:clustering algorithm} with the determined $h$ values results in a satisfying clustering of particles that is visualized in the particle clustering maps in Appendix~\ref{appendix_dust_clustering_maps}. 

\begin{table}[]
\caption{Mean particle velocities measured by GIADA, the applied bandwidth value~$h$ in the mean-shift algorithm and corresponding cluster sizes.} 
\label{table:band_width_value_cluster}
\renewcommand{\arraystretch}{1.2}
\centering \begin{tabular}{llll}
\toprule
Target & \begin{tabular}[c]{@{}l@{}}Target  \\ mean \\ velocity $v$\\ {[}m s$^{\rm -1}${]}\end{tabular} & \begin{tabular}[c]{@{}l@{}}Applied          \\ band \\width $h$\end{tabular} & \begin{tabular}[c]{@{}l@{}}corresponding    \\ cluster radius \\ {[}$\mu$m{]}\end{tabular} \\\hline
11     &     3.1 $\pm$ 1.5          & 14     & $\sim$25     \\
13     &     2.7 $\pm$ 0.3          & 14     & $\sim$25     \\
14     &     7.21 $\pm$ 0.09        & 24     & $\sim$35     \\
15     &     7.2 $\pm$ 0.2          & 24     & $\sim$35     \\ 
\bottomrule
\end{tabular}
\end{table}

Table~\ref{table:number_of_cluster} shows a summary of the number of particles and parent particles (clusters) per target together with their ratio. We find that the average ratio between fragments and parent particles (clusters) is $\sim$ ~4.1 much smaller for targets 11 and 13 (and 15, however, the statistics of target 15 are negligible) than with $\sim$ 11.8 for target 14. In particular, a comparison of the circumstances of dust collection showed that particles on targets 11 and 13 were collected during nominal comet activity pre-perihelion in late 2014 and early 2015 with lower collection velocities, but particles on target 14 were gathered during a post-perihelion outburst on February 2016 with some meters per second higher impact speed.

Furthermore, the clustering results achieved with this algorithm are assumed to be reasonable as they match with the results from complementary studies~(\citealp{Ellerbroek_labstudy_2017, Longobardo_2022_MNRAS}). In particular, \citet{Longobardo_2022_MNRAS} carried out a complementary approach to derive MIDAS cluster sizes: assuming that dust particles clustered within two times the fragment's size (lying on the same target and detected in the same exposure period) belong to the same parent particle, they find similar results concerning the cluster sizes and fragmentation ratios. In agreement with~\citet{Longobardo_2022_MNRAS}, we find that the parent particles collected during the outburst are easier fragmented than those collected during nominal comet activity pre-perihelion. We suggest that the reason is not only driven by the higher impact speed but also higher particle compaction with lower deposition speed (see Section \ref{sec:results_Basic_statistics}; \citealp{Longobardo_2022_MNRAS}). We would like to open up the possibility that the difference in density detected by GIADA (\citealp{Longobardo_2022_MNRAS}) is a driver of higher fragmentation ratios, where lower density particles are more prone to fragmentation either due to higher porosity or different material composition. 

\begin{table}[]
\caption{Number of particles and parent particles (clusters), and MIDAS fragmentation ratio for each target.} 
\label{table:number_of_cluster}
\renewcommand{\arraystretch}{1.2}
\centering \begin{tabular}{llll}
\toprule
Target & \begin{tabular}[c]{@{}l@{}}Number of\\particles\end{tabular} & \begin{tabular}[c]{@{}l@{}}Number of\\parent particles\\(cluster)\end{tabular} & \begin{tabular}[c]{@{}l@{}}MIDAS\\fragmentation \\ ratio\end{tabular}\\\hline
11      & 350       & 55        & 6.36      \\
13      & 61        & 15        & 4.1       \\
14      & 1471      & 125       & 11.8      \\
15      & 7         & 4         & 1.8       \\                             
\bottomrule
\end{tabular}
\end{table}


\section{Summary}\label{sec:conclusions}
        
The Micro-Imaging Dust Analysis System (MIDAS) atomic force microscope on board the Rosetta comet orbiter investigated and measured the 3D topography of a few hundred of nanometer- to tens of micrometer-sized cometary dust particles with resolutions down to a few nanometers. We updated and enhanced a MIDAS particle catalog of all detected MIDAS dust particles together with their main properties such as size, height, basic shape descriptors, and collection time. We present scientific results directly extracted from the catalog such as the size distribution and statistical characteristics of cometary dust particles. 

The majority of the MIDAS particles are assumed to be fragmented upon collection. This is not only supported by the herein presented clustering maps but also suspected through observations of the two complementary dust analysis instruments COSIMA and GIADA on board Rosetta. Thus, the following dust particle sizes and size distributions may be heavily influenced by fragmentation during dust collection and are not easily comparable to unaltered particles in the cometary coma. The key results are as follows:

\begin{itemize}
    \item We updated the MIDAS particle catalog to version V6 to contain all dust particles detected by MIDAS and displayed them as a table with a row for each particle and columns for the (meta) data. The new version of the MIDAS catalog includes 3523 particles in total. We carefully selected 2088 particles (493 particles before perihelion + 1595 particles after perihelion), whose image quality is high enough that they can be used in further studies and selected the best representation for each multiply scanned particle, which left us with 1857 particles for the presented studies. \newline

    \item We determined the equivalent radius based on the 2D projection of MIDAS particles to range from about 80~nm to about 8.5~$\mu$m in size. In particular, the mean value of particle equivalent radius on the 2D projection is 0.91~$\pm$~0.79~$\mu$m. As an improvement to the commonly used equivalent radius, we developed an equivalent radius that takes into account the 3D nature of the MIDAS data set, namely an equivalent radius based on the volume measurement of the particles. This volume-derived equivalent radius with a mean value of 0.56~$\pm$~0.45~$\mu$m does agree with the 2D projection-based equivalent radius within the uncertainty. \newline

    \item We compared the equivalent radii of particles collected pre-perihelion to those radii derived for particles from post-perihelion and found that the calculated values themselves do not show a noticeable difference. This implies that particles have similar projected areas and volume regardless of cometary activities, leading to the conclusion that different cometary activities (e.g., nominal activity or outburst) do not relate to either particle sizes or the degree of compression.\newline
    
    \item We compared the measured volume of MIDAS particles to the calculated volumes of commonly found shapes such as cone shapes and cylinder shapes. A comparison of the derived volume-based equivalent radius suggested that the average MIDAS particle shape looks like a cone shape but with rounded tips.\newline 

    \item We determined the size distribution for the two types of equivalent radii. We found that the slope of the size distribution of the equivalent radius based on the 2D projection shows a similar behavior to the one derived from volume. Both follow a power law ${r}^{\rm{\,b}}$ fit with an index $\rm{\,b}$ of $\sim$ -1.67 $\pm$ 0.11 to -1.88 $\pm$ 0.04, respectively.
    The slight differences in slope for the two equivalent radii may reflect lower compaction of small fragments: the projected area of the particles shrinks faster than the related volume of the particles. There seems to be a population of less compressed small particles, potentially fragments that avoided compression because the impact energy was consumed by fragmentation or different material composition. \newline
    
    \item The relative similarity of the slopes of the size distributions for each period (with an index of -1.23~$\pm$~0.10 to -1.28~$\pm$~0.08 for outburst activity in post-perihelion versus an index of -1.05~$\pm$~0.10 to -1.19~$\pm$~0.08 for nominal activity in pre-perihelion) may imply that different cometary activities and impact velocity may not play the driving role in the fragmentation of larger particles, but only in the number of collected parent particles.\newline
    
    \item We investigated fragmentation mechanisms during dust collection with the goal of determining the arriving MIDAS parent particle sizes. We found indications for substantial loss of parent particle mass upon collection and thus could only roughly estimate MIDAS parent particle sizes to be in the tens of micrometer range. The created cluster sizes by fragmentation of such a parent particle were roughly approximated to be about 25 to 35~$\mu$m in radius depending on the impact velocity. The slope of the size distribution for the MIDAS parent particle size is $\sim$ 1.44, showing consistency with the finding from COSIMA for smaller particles. \newline

    \item Based on the newly updated MIDAS particle catalog, we created dust coverage maps of the MIDAS collection targets to obtain an easily accessible overview of dust deposition. 
    We ran a sophisticated mean-shift algorithm to determine clusters of particles on the dust coverage maps. On this basis, we created dust clustering maps to trace possible fragmentation of collected particles. We determined the fragmentation ratio, namely, how many fragments were created per parent particle, to be 4.09 for nominal activity and 11.8 for the February 16, 2016 outburst. With the help of GIADA data about particle velocities and densities, we could determine that parent particles seem to undergo heavier fragmentation if they have higher deposition velocities and/or lower densities. \newline

\end{itemize}

\noindent The comprehensive MIDAS catalog is a unique and useful tool for identifying and investigating micrometer-sized cometary dust particles, allowing for quick access to a variety of particle characteristics, properties, and statistics, and it has been made available and updated in the ESA/PSA. This catalog can be the basis for all following scientific investigations, for instance, the development of more sophisticated MIDAS particle shape descriptors.


\begin{acknowledgements}
We deeply miss our passed away MIDAS Co-I Prof. Anny-Chantal Levasseur-Regourd. We were grateful to have her in our team as her energy and enthusiasm truly inspired us. The authors gratefully thank our referee for the constructive comments and recommendations which definitely help to improve the readability and quality of the paper. M. Kim and T. Mannel acknowledge funding by the ESA project "Primitiveness of cometary dust collected by MIDAS on-board Rosetta" (Contract No. 4000129476). This research was supported by the International Space Science Institute (ISSI) through the ISSI International Team “Characterization of cometary activity of 67P/Churyumov-Gerasimenko comet”.
\end{acknowledgements}

\bibliographystyle{aa} 
\bibliography{AA_2022_45262.bib} 

\begin{appendix}

\section{Terminology}\label{sec:terminology}

Despite the complexity of cometary dust structures, a dust classification scheme has already been unified with a clearly defined vocabulary in the Rosetta dust studies (\citealp{Mannel_classification_2019, Guettler_morphology_2019}). Here, we present the terminology for all the different components of dust referred to in this study and as they apply to collection and detection by MIDAS, in particular:

\begin{itemize}

\item \textbf{Particle}:  a subordinate term that describes any type of dust unit (grain, fragment, subunit, and/or cluster). If the type of dust is not known (e.g., whether it is a fragment), we suggest using this term. \newline

\item \textbf{Fragment}:  indicates that a larger particle was broken into pieces. A fragment can be pre-existing in the dust collection (e.g., as a natural subunit formed in the dust growth processes) or created by impact and/or other processes. \newline

\item \textbf{Parent particle}:  describes a particle as it enters the MIDAS instrument before hitting the collection target. The parent particle can either stay intact (i.e, unaltered by collection) or be altered (e.g., compression, fragmentation). \newline

\item \textbf{Cluster}: Cluster indicates a collection of several fragments spatially so close to each other that they are expected to originate from the same parent particle. Clusters always consist of fragments, which can also be called particles because the particle is a subordinate term. The fragments may contain subunits, which are either clusters of grains or grains. \newline

\item \textbf{Grain}:  the smallest and fundamental building blocks. They need to be morphologically closed entities but can be chemically diverse (e.g., hard spheres). \newline 

\item \textbf{Grain cluster}: describes a conglomerate of grains. It may be possibly grown by hierarchical patterns in the early Solar System. \newline

\item \textbf{Subunit}:  indicates a morphologically distinct part in a particle. It can either be a grain or a conglomerate of grains. In some cases, differentiation is not possible due to low resolution or imaging artifacts. \newline
    
\end{itemize}

\onecolumn
\section{Target exposure history}\label{appendix: Target exposure history}
From the start of the Rosetta mission, 15 targets were exposed in total. We note that this does not necessitate that any dust is collected at this time. However, only 8 targets were exposed and scanned for dust collection after the arrival of the Rosetta spacecraft at comet 67P and only 4 targets (i.e., targets 11, 13, 14, and 15) were found to have detectable dust deposits of cometary origin. The full table of the MIDAS target exposure history is available at the CDS.\newline

\tablefirsthead{\toprule Start & End & Duration & Target & Note \\ \midrule}
\tablehead{%
\multicolumn{5}{c}%
{{}} \\
\toprule Start & End & Duration & Target & Note\\ \midrule}
\tabletail{%
\midrule \multicolumn{5}{r}{{}} \\ }
\tablelasttail{%
\\}
\renewcommand{\arraystretch}{1.2}
\tablecaption{Target exposure history with start and end time, and duration, for each target with comet activities.}
\begin{minipage}{\linewidth}
\begin{supertabular}{lllll}
\label{table:target exposure history}
2004-04-04 22:58:07 & 2004-04-04 23:04:38 & 0 days 00:06:31    & 33     &      \\
2005-03-27 00:35:15 & 2005-03-27 00:38:13 & 0 days 00:02:58    & 33     &      \\
2005-04-19 13:03:24 & 2005-04-19 13:06:23 & 0 days 00:02:58    & 33     &      \\
2005-10-05 00:42:16 & 2005-10-05 00:44:14 & 0 days 00:01:57    & 33     &      \\
2006-03-04 19:00:17 & 2006-03-04 19:03:16 & 0 days 00:02:58    & 33     &      \\
2006-03-08 18:00:17 & 2006-03-08 18:03:15 & 0 days 00:02:58    & 33     &      \\
2006-08-26 18:00:15 & 2006-08-26 18:03:13 & 0 days 00:02:57    & 33     &      \\
2006-08-30 17:00:15 & 2006-08-30 17:03:13 & 0 days 00:02:57    & 33     &      \\
2006-11-23 10:00:17 & 2006-11-23 10:03:15 & 0 days 00:02:57    & 33     &      \\
2007-05-19 14:00:16 & 2007-05-19 14:03:14 & 0 days 00:02:58    & 34     &      \\
2007-05-23 13:00:16 & 2007-05-23 13:03:14 & 0 days 00:02:58    & 33     &      \\
2007-09-16 13:00:18 & 2007-09-16 13:03:16 & 0 days 00:02:58    & 33     &      \\
2007-10-04 19:02:17 & 2007-10-04 19:07:15 & 0 days 00:04:58    & 34     &      \\
2008-01-05 12:34:14 & 2008-01-05 12:37:12 & 0 days 00:02:57    & 2      &      \\
2008-01-09 14:34:15 & 2008-01-09 14:37:11 & 0 days 00:02:55    & 2      &      \\
2008-07-05 04:15:17 & 2008-07-05 04:18:17 & 0 days 00:02:59    & 33     &      \\
2009-01-29 09:51:17 & 2009-01-29 09:54:16 & 0 days 00:02:58    & 3      &      \\
2009-02-02 15:51:18 & 2009-02-02 15:54:16 & 0 days 00:02:58    & 33     &      \\
2009-09-20 17:45:17 & 2009-09-20 17:47:17 & 0 days 00:02:00    & 33     &      \\
2010-04-24 21:15:17 & 2010-04-24 21:17:17 & 0 days 00:02:00    & 35     &      \\
2010-07-10 11:09:19 & 2010-07-10 22:51:18 & 0 days 11:41:58    & 37     &      \\
2010-12-03 23:00:20 & 2010-12-03 23:02:18 & 0 days 00:01:58    & 5      &      \\
2010-12-07 00:30:20 & 2010-12-07 00:32:18 & 0 days 00:01:58    & 33     &      \\
2010-12-07 02:25:21 & 2011-03-05 01:02:19 & 87 days 22:36:58   & 33     &      \\
2011-03-05 02:55:21 & 2014-09-03 21:58:18 & 1278 days 19:02:57 & 33     &      \\
2014-09-17 12:27:20 & 2014-09-21 12:15:21 & 3 days 23:48:00    & 11     &      \\
2014-09-28 00:06:20 & 2014-09-29 10:35:22 & 1 days 10:29:02    & 11     &      \\
2014-09-30 12:16:21 & 2014-10-01 13:17:22 & 1 days 01:01:01    & 11     &      \\
\multicolumn{5}{c}{...} \\
2016-02-16 23:25:25 & 2016-02-20 05:55:19 & 3 days 06:29:54    & 14     &  Outburst    \\
2016-02-21 08:07:26 & 2016-02-22 21:08:20 & 1 days 13:00:54    & 14     &      \\
2016-02-23 23:19:26 & 2016-02-24 04:55:01 & 0 days 05:35:35    & 14     &      \\
2016-03-03 11:18:17 & 2016-03-05 07:23:31 & 1 days 20:05:14    & 12     &      \\
2016-03-08 10:50:38 & 2016-03-09 05:57:11 & 0 days 19:06:33    & 12     &      \\
\hline\multicolumn{5}{c}{...\footnote{The full table is available at the CDS.}}\\
\bottomrule
\end{supertabular}
\end{minipage}

\section{Best representation of each particle}\label{appendix: Best representation of each particle}

The MIDAS particle catalog presented in this paper contains in total of 3523 scans of particles. Since a part of those scans are suffering from artifacts, low resolution, or represent the same particle just with different scanning parameters, there was a cautious selection of particles used for further scientific analysis necessary. In Sect.~\ref{sec:results_Basic_statistics}, the steps of this selection process are described. The full table of the ultimately chosen 1857 particles is available at the CDS.\newline 

\begin{landscape}
\begin{table}
\caption{Scanning time, 2D and 3D radius, location of dust particles, and the cluster number (i.e., a reference to the assumed cluster or parent particle family) of best representation of each particle.}
\begin{minipage}{\linewidth}
\scriptsize
\begin{tabular}{llllllllllll}

\toprule \renewcommand{\arraystretch}{1.2}

\multirow{3}{*}{\textbf{Particle ID}} & \textbf{Target} & \textbf{Start time} & \textbf{Duration} & \textbf{End time} & \textbf{2D radius\footnote{the equivalent radius based on the 2D projection}} & \textbf{3D radius\footnote{the equivalent radius derived from volume}} & \multicolumn{4}{c}{\textbf{Location}\footnote{from the center of the individual target}}  & \textbf{Cluster} \\ 
& \multirow{2}{*}{\textbf{Number}} & \multirow{2}{*}{Y-M-D H:M:S} & \multirow{2}{*}{Days H:M:S}& \multirow{2}{*}{Y-M-D H:M:S} & \multirow{2}{*}{[$\mu$m]} & \multirow{2}{*}{[$\mu$m]} & \textbf{Left} & \textbf{Right} & \textbf{Up} & \textbf{Down}& \multirow{2}{*}{\textbf{Number}} \\ 
&  &  & &  &  & & [$\mu$m] & [$\mu$m] & [$\mu$m] & [$\mu$m] &  \\ \hline
2014-11-18T032520\_P01\_T11 & 11 & 2014-11-18 03:25:20.311766 & 0 days 06:48:49.008102 & 2014-11-18 10:14:09.319868 & 0.80 & 0.51 & -130.20 & -128.52 & 28.11 & 26.85 & 49 \\ 
    2014-11-18T131924\_P01\_T11 & 11 & 2014-11-18 13:19:24.323539 & 0 days 06:53:45.008200 & 2014-11-18 20:13:09.331739 & 0.46 & 0.26 & 63.31 & 64.15 & 53.70 & 52.65 & 11 \\ 
    2014-11-18T131924\_P02\_T11 & 11 & 2014-11-18 13:19:24.323539 & 0 days 06:53:45.008200 & 2014-11-18 20:13:09.331739 & 0.94 & 0.67 & 63.31 & 65.40 & 57.06 & 54.96 & 11 \\ 
    2014-11-18T131924\_P03\_T11 & 11 & 2014-11-18 13:19:24.323539 & 0 days 06:53:45.008200 & 2014-11-18 20:13:09.331739 & 0.34 & 0.18 & 65.82 & 66.45 & 52.23 & 51.60 & 11 \\ 
    2014-11-18T131924\_P04\_T11 & 11 & 2014-11-18 13:19:24.323539 & 0 days 06:53:45.008200 & 2014-11-18 20:13:09.331739 & 0.36 & 0.21 & 35.83 & 36.67 & 40.49 & 39.86 & 44 \\ 
    2014-11-18T131924\_P05\_T11 & 11 & 2014-11-18 13:19:24.323539 & 0 days 06:53:45.008200 & 2014-11-18 20:13:09.331739 & 0.62 & 0.40 & 72.75 & 73.80 & 30.00 & 28.53 & 8 \\ 
    2014-11-18T131924\_P06\_T11 & 11 & 2014-11-18 13:19:24.323539 & 0 days 06:53:45.008200 & 2014-11-18 20:13:09.331739 & 0.84 & 0.52 & 73.17 & 74.84 & 28.11 & 26.22 & 8 \\ 
    2014-11-18T131924\_P07\_T11 & 11 & 2014-11-18 13:19:24.323539 & 0 days 06:53:45.008200 & 2014-11-18 20:13:09.331739 & 0.25 & 0.13 & 69.18 & 69.60 & 41.33 & 40.91 & 11 \\ 
    2014-11-18T131924\_P08\_T11 & 11 & 2014-11-18 13:19:24.323539 & 0 days 06:53:45.008200 & 2014-11-18 20:13:09.331739 & 0.18 & 0.15 & 67.29 & 67.50 & 49.09 & 48.67 & 11 \\ 
    2014-11-18T131924\_P09\_T11 & 11 & 2014-11-18 13:19:24.323539 & 0 days 06:53:45.008200 & 2014-11-18 20:13:09.331739 & 0.22 & 0.15 & 72.75 & 72.96 & 31.89 & 31.26 & 8 \\ 
    2014-11-18T131924\_P10\_T11 & 11 & 2014-11-18 13:19:24.323539 & 0 days 06:53:45.008200 & 2014-11-18 20:13:09.331739 & 0.31 & 0.17 & 63.52 & 64.15 & 54.33 & 53.70 & 11 \\ 
    2014-11-18T131924\_P11\_T11 & 11 & 2014-11-18 13:19:24.323539 & 0 days 06:53:45.008200 & 2014-11-18 20:13:09.331739 & 0.18 & 0.13 & 63.10 & 63.31 & 54.33 & 53.91 & 11 \\ 
    2014-11-27T193208\_P02\_T11 & 11 & 2014-11-27 19:32:08.587771 & 0 days 07:12:05.008563 & 2014-11-28 02:44:13.596334 & 0.21 & 0.15 & -16.91 & -16.43 & 2.39 & 1.98 & 46 \\ 
    2014-11-28T034542\_P02\_T11 & 11 & 2014-11-28 03:45:42.597553 & 0 days 07:08:14.008487 & 2014-11-28 10:53:56.606040 & 0.25 & 0.14 & -3.70 & -3.15 & 12.35 & 11.87 & 33 \\ 
    2014-11-28T034542\_P03\_T11 & 11 & 2014-11-28 03:45:42.597553 & 0 days 07:08:14.008487 & 2014-11-28 10:53:56.606040 & 0.22 & 0.14 & -0.13 & 0.42 & 14.89 & 14.47 & 33 \\ 
    2014-11-29T222402\_P06\_T11 & 11 & 2014-11-29 22:24:02.648255 & 0 days 13:54:45.016543 & 2014-11-30 12:18:47.664798 & 0.32 & 0.18 & -79.03 & -78.19 & 19.20 & 18.66 & 13 \\ 
    2014-11-29T222402\_P07\_T11 & 11 & 2014-11-29 22:24:02.648255 & 0 days 13:54:45.016543 & 2014-11-30 12:18:47.664798 & 0.27 & 0.19 & -76.00 & -75.46 & 24.96 & 24.41 & 13 \\ 
    2014-11-29T222402\_P08\_T11 & 11 & 2014-11-29 22:24:02.648255 & 0 days 13:54:45.016543 & 2014-11-30 12:18:47.664798 & 0.13 & 0.08 & -75.86 & -75.46 & 24.16 & 23.96 & 13 \\ 
    2014-11-29T222402\_P09\_T11 & 11 & 2014-11-29 22:24:02.648255 & 0 days 13:54:45.016543 & 2014-11-30 12:18:47.664798 & 0.13 & 0.08 & -75.81 & -75.41 & 24.16 & 23.96 & 13 \\ 
    2014-11-29T222402\_P10\_T11 & 11 & 2014-11-29 22:24:02.648255 & 0 days 13:54:45.016543 & 2014-11-30 12:18:47.664798 & 0.16 & 0.11 & -76.50 & -76.20 & 24.81 & 24.41 & 13 \\ 
    2014-11-29T222402\_P11\_T11 & 11 & 2014-11-29 22:24:02.648255 & 0 days 13:54:45.016543 & 2014-11-30 12:18:47.664798 & 0.29 & 0.19 & -72.43 & -71.79 & 32.44 & 31.75 & 13 \\ 
    2014-11-30T141103\_P02\_T11 & 11 & 2014-11-30 14:11:03.667023 & 0 days 13:50:28.016458 & 2014-12-01 04:01:31.683481 & 0.14 & 0.09 & -67.28 & -66.98 & 26.44 & 26.15 & 13 \\ \hline
    \multicolumn{12}{c}{...\footnote{The full table is available at the CDS.}} \\
\bottomrule 

\end{tabular}
\end{minipage}
\label{table:Best representation of 1857 particles}

\end{table}

\end{landscape}

\twocolumn

\section{Estimating the value of bandwidth $h$}\label{appendix: band_width}

\begin{figure*}
\centering
\includegraphics[width=13cm, height=17cm]{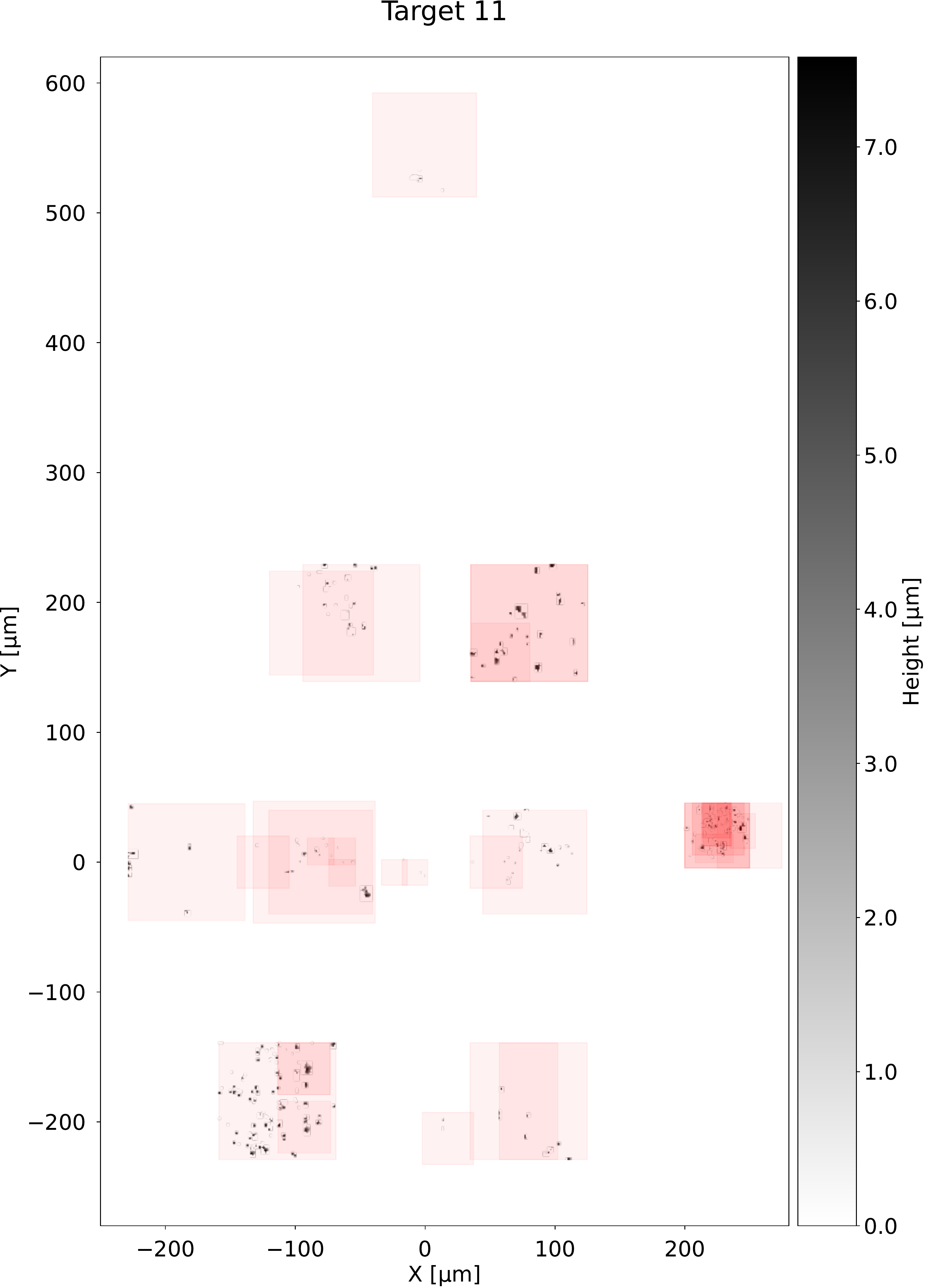}
\caption{2D dust coverage map of target 11 only including fully imaged particles as selected in Section~\ref{sec:results_Basic_statistics} with an indicated scanned area (red boxes; overlaying scans create darker red color, thus more intense color indicates iterative scanning).}
\label{fig:2D_dust_coverage_map_10}
\end{figure*}

Estimating the value of scaling factor bandwidth, $h,$ is significantly important as choosing this value crucially impacts the cluster size and interpretation. We calculated the estimated value of bandwidth, $h,$ in the mean shift method with the k-nearest neighbors (kNN) algorithm from \texttt{estimate\_bandwidth}\,\footnote{\url{https://scikit-learn.org/stable/modules/generated/sklearn.cluster.estimate_bandwidth.html}} function in Python module \texttt{sklearn.cluster},  evaluating the isolation and connectivity of the delineated clusters. A non-parametric KNN algorithm (\citealp{Fix1989, Cover1967}) relies on observable data similarities and sophisticated distance metrics (e.g., Euclidean, Manhattan distance, or Minkowski Distance; default and standard choice is the Euclidean distance) from new unclassified or unlabeled data points to existing classified or labeled data points to generate accurate predictions of the value of bandwidth, $h,$ in the mean shift method (\citealp{Comaniciu2002}). The distance metrics can be described as: 

\begin{equation}
{\rm{Euclidean\,distance}}\,(x,\,y) = \bigg(\,\sum_{i = 1}^{n} \left|{x_{\rm{\,i}} - y_{\rm{\,i}}} \right|^{p}\bigg)^{\frac{1}{p}}: p = 2,
\label{eq: distance}
\end{equation} 

where $n$ is the number of dimensions (2 for our analysis), $x$ is the data point from the dataset, and $y$ is the new data point to be predicted. Since the kNN is a type of classification where the function is approximated locally and MIDAS particle detection is scanning region-dependent (e.g., see Figure~\ref{fig:2D_dust_coverage_map_10}), we thus divided MIDAS particle detection in target 11 into locally ($\sim$ 100 $\times$ 100 ${\mu}$m), namely, regions 1 to 8 (see Table \ref{table:bandwidth and Silhouette} for details). Estimated bandwidth values from \texttt{estimate\_bandwidth} function and the corresponding number of clusters in the local regions can be found in Table \ref{table:bandwidth and Silhouette}.

The bandwidth, $h,$ and thus cluster number can be also validated by an alternative method such as the silhouette analysis (\citealp{Rousseeuw1987}). This method estimates the average distance between clusters' so-called silhouette coefficient $S_{\rm i}$ (also: silhouette width or score) for each observation $i$ and a measure of how close each point in one cluster is to points in the neighboring clusters, validating the optimal number of clusters. Even though the silhouette coefficient mostly used the clustering algorithm to configure the number of clusters in advance, the optimal cluster number from this method could be compared to the one from the result of the mean-shift method with the estimated value of bandwidth $h$ by the kNN algorithm. The silhouette coefficient, $S_{\rm{i}}$, can be described as follows:

\begin{equation}
S_{\rm i} = \frac{b_{\rm\,i}-a_{\rm\,i}}{\rm{max}\,(a_{\rm\,i},\,b_{\rm\,i})},
\label{eq: Silhouette}
\end{equation} 

where $a_{\,\rm i}$ is the average dissimilarity between $i$ and all other points of the cluster to which $i$ belongs and $b_{\,\rm i}$ is the smallest of $d\,(i,\,C)$ for all other clusters $C$ to which $i$ does not belong. Its value ranges from \,-1 (i.e., clusters are assigned in the wrong way) to 1 (i.e., clusters are well apart from each other and clearly distinguished). The optimal cluster numbers from the silhouette analysis can be found in Table \ref{table:bandwidth and Silhouette}. 

\begin{figure*}
\includegraphics[width=12cm, height=8cm]{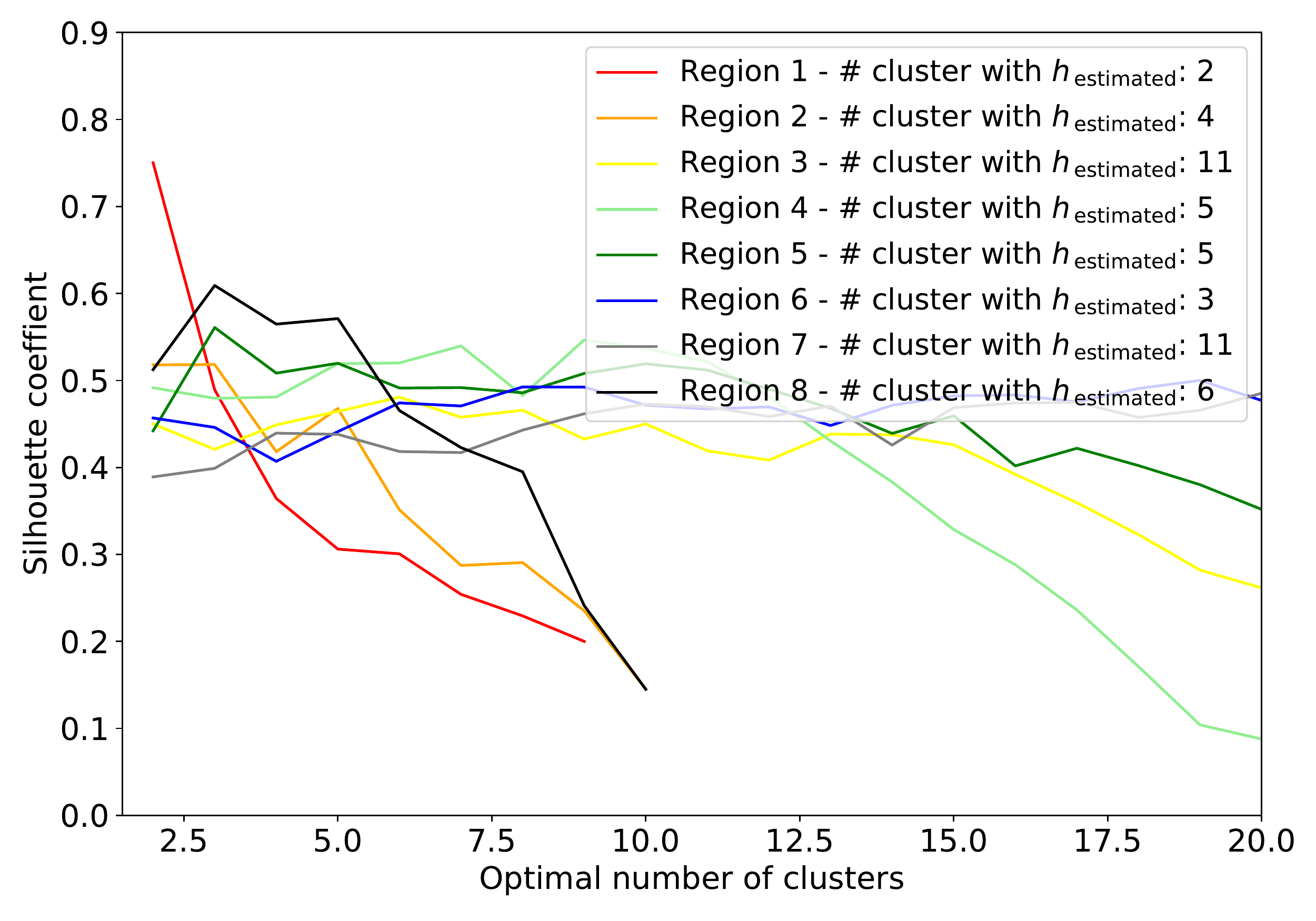}
\caption{Result from the silhouette coefficient analysis to find an optimal number of clusters. The corresponding number of clusters derived from the bandwidth, $h,$ is listed in the legend and Table \ref{table:bandwidth and Silhouette}, which shows consistency with the optimal number of clusters with the highest silhouette coefficient for the individual areas (see Table \ref{table:bandwidth and Silhouette}).}
\label{fig:silhouette coefficient}
\end{figure*}

We find that the optimal cluster numbers that are calculated by the k-nearest neighbors algorithm and the silhouette coefficient analysis match well in a reasonable range (at least with the local maxima of the silhouette coefficient; see Table \ref{table:bandwidth and Silhouette} and Figure \ref{fig:silhouette coefficient}).

Finally, we compared our results to the random distribution of particles (e.g., the same number of particles within the same local area) to see if all MIDAS dust fragments are in reality fragments of parent particles that are not randomly distributed over the target. We used the same bandwidth (i.e., the same size of Kernel distribution) and checked that the results vary substantially. In particular, the result with the random distribution underestimates the number of clusters. Thus, we concluded that our cluster interpretation is more statistically significant than the random distribution interpretation.

Even though these methods provide metrics calculating the goodness of a clustering technique, we mainly investigate a more qualitative inspection (e.g., comparing the cluster size that is derived from the bandwidth, $h,$ by the kNN algorithm and the number of clusters from silhouette coefficient to the previous experimental study of \citealp{Ellerbroek_labstudy_2017}) and chose the one that looks the most reasonable (see also Section \ref{sec:clustering algorithm} and \ref{sec:The_size_of_MIDAS_clusters}). Therefore, we decide to use 14 as a bandwidth, $h,$ for the case of Target 11.

\begin{table*}[]
\renewcommand{\arraystretch}{1.2}
\centering \begin{tabular}{lllll}
\hline\hline 
Region& Location from the center& Estimated bandwidth& Corresponding& Optimal number of cluster\\ 
Number& [$\mu$m]& $h_{\rm{\,estimated}}$& number of cluster& from the silhouette coefficient\\ \hline
Region 1& -50 < X < 50, 500 < Y < 600& 3.55& 2& $\sim$ 2\\
Region 2& -100 < X < 0, 150 < Y < 250& 9.14& 4& $\sim$ 3\\
Region 3& 0 < X < 150, 100 < Y < 250& 24.94& 11& $\sim$ 10\\
Region 4& -150 < X < -50, -50 < Y < 50& 14.32& 5& $\sim$ 5-7\\
Region 5& 50 < X < 150, -50 < Y < 50& 14.54& 5& $\sim$ 3\\
Region 6& 200 < X < 300, 0 < Y < 100& 11.57& 3& $\sim$ 3, 8-10\\
Region 7& -150 < X < -50, 0 <  Y < -100& 26.44& 11& $\sim$ 11-13\\
Region 8& 0 < X < 100, -250 < Y < -150& 9.52& 6& $\sim$ 4\\ \hline
& Mean $\pm$ std &  14.25 $\pm$ 7.36& &\\ \hline
\end{tabular}
\caption{ Estimated value of bandwidth and the corresponding number of clusters for Target 11 with the k-nearest neighbors algorithm and silhouette coefficient analysis.}
\label{table:bandwidth and Silhouette}
\end{table*}

\section{Visualization of the MIDAS dust coverage maps}\label{appendix_dust_coverage_maps}

Overall, MIDAS dust coverage maps are 2D and 3D grayscale images showing the exact particle topographies at their respective locations on the target. As described in Section~\ref{sec:results_coverage_map} a variety of maps were created and in the following the most important ones for good visualization of the MIDAS data set are shown: the 2D dust coverage maps including only the fully imaged particles usable for further analysis (as selected in Section~\ref{sec:results_Basic_statistics}) with indicated scanning area, and the 3D dust coverage maps including the same particles. We note that we explicitly exaggerate the height of particles (i.e., Z plane) compared to the X-Y plane in the 3D dust coverage maps to show the topographies of MIDAS particles clearly. 

\subsection{2D MIDAS dust coverage maps}\label{appendix_2D_dust_coverage_maps}

\begin{figure*}
\centering
\includegraphics[width=11cm, height=17cm]{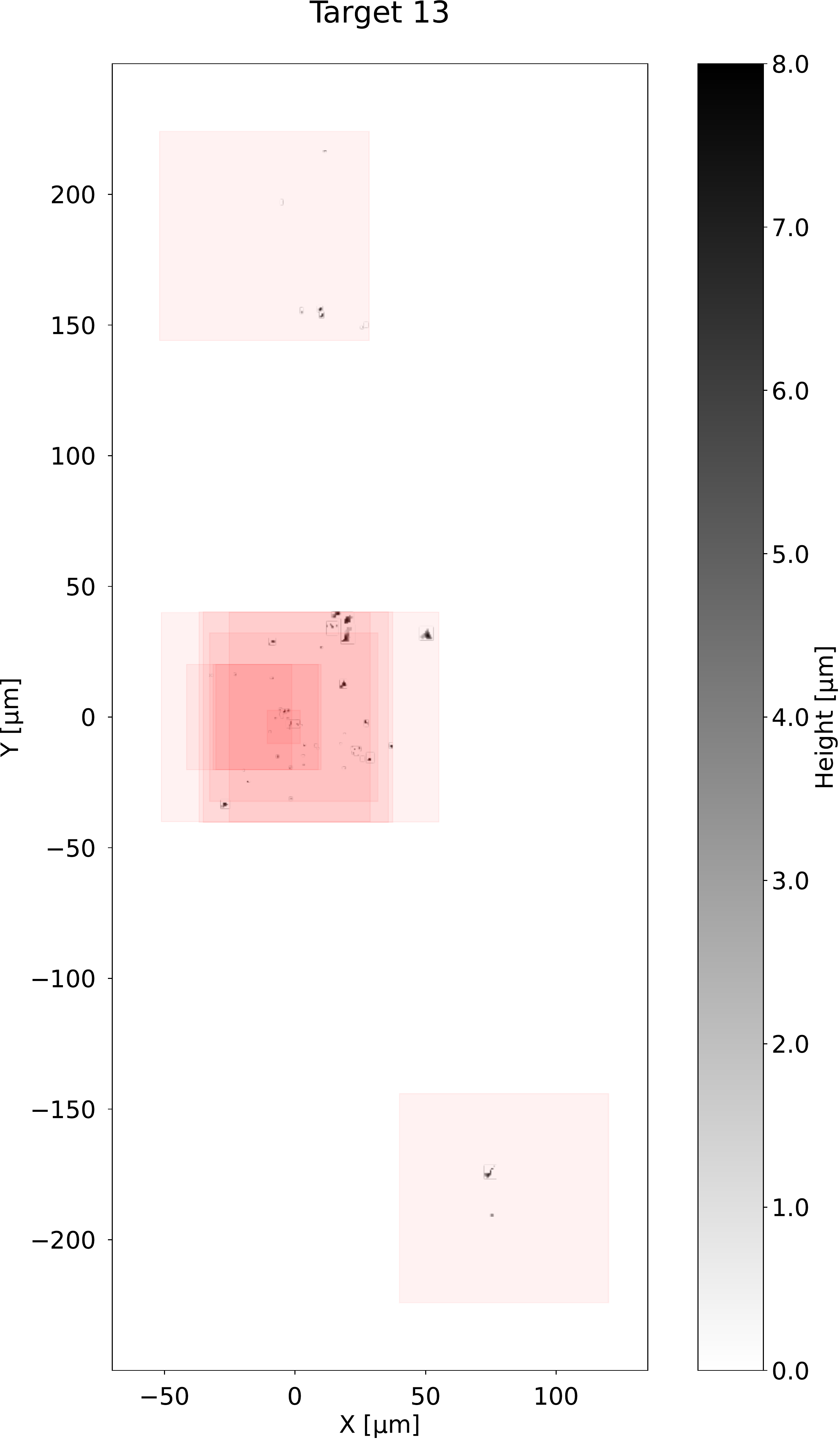}
\caption{2D dust coverage map of target 13 only including fully imaged particles as selected in Section~\ref{sec:results_Basic_statistics} with an indicated scanned area (red boxes; the darkness of red color indicates iterative scanning).}
\label{fig:2D_dust_coverage_map_12}
\end{figure*}

\begin{figure*}
\centering
\includegraphics[width=15cm, height=24cm]{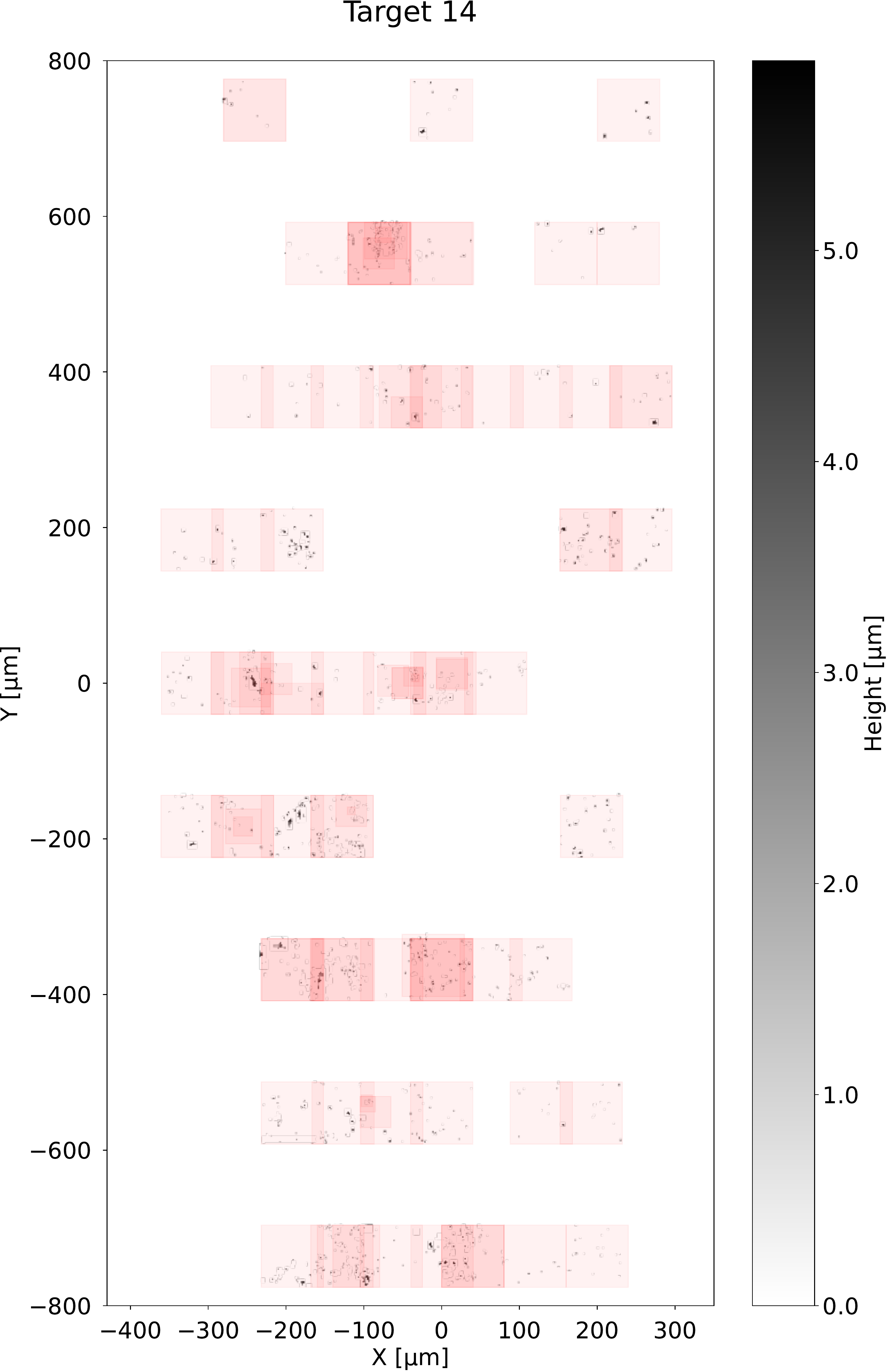}
\caption{2D dust coverage map of target 14 only including fully imaged particles as selected in Section~\ref{sec:results_Basic_statistics} with an indicated scanned area (red boxes; the darkness of red color indicates iterative scanning).}
\label{fig:2D_dust_coverage_map_13}
\end{figure*}

\begin{figure*}
\centering
\includegraphics[width=12 cm, height=9cm]{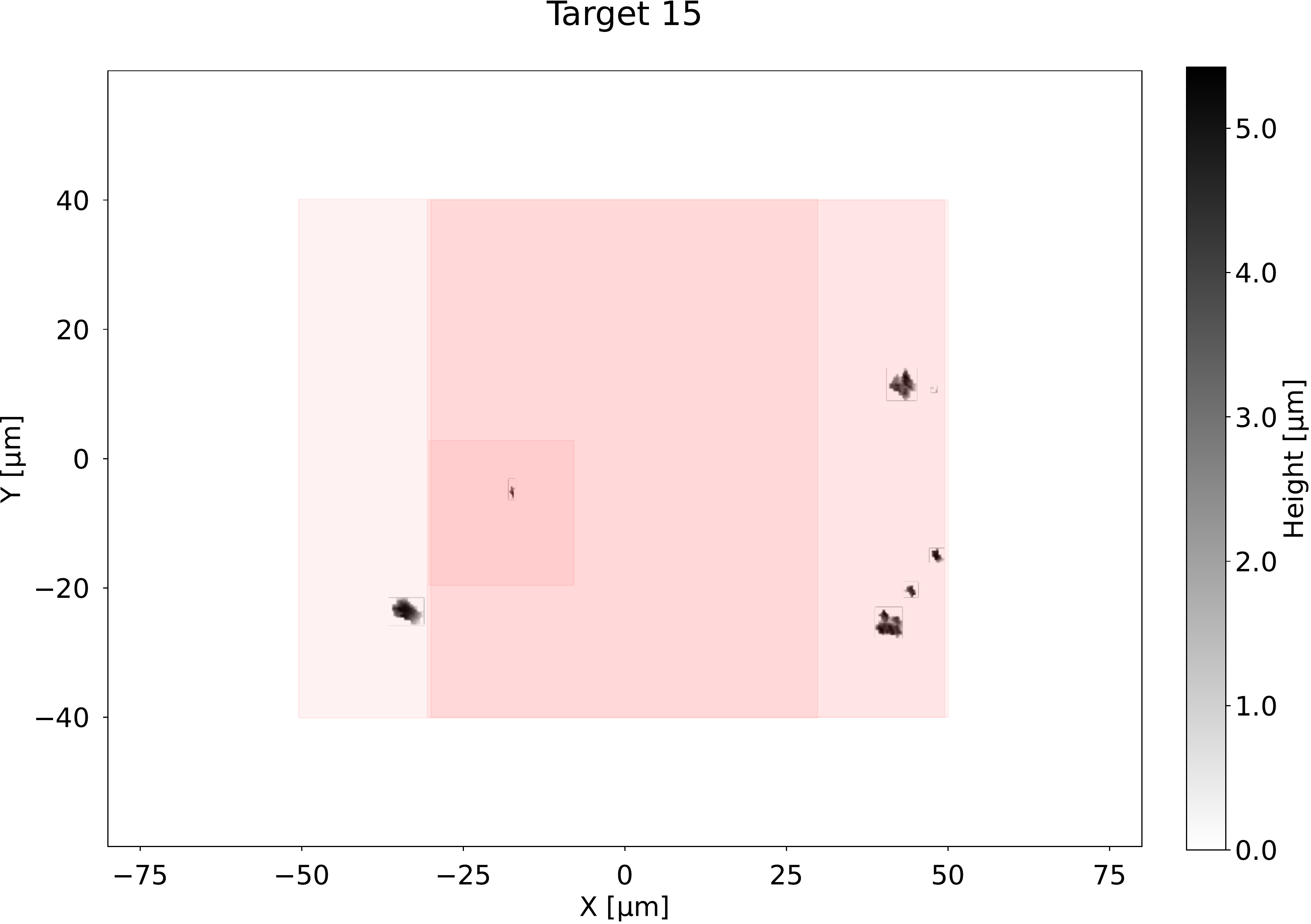}
\caption{2D dust coverage map of target 15 with fully imaged particles as selected in Section~\ref{sec:results_Basic_statistics} with an indicated scanned area (red boxes; the darkness of red color indicates iterative scanning.}
\label{fig:2D_dust_coverage_map_14}
\end{figure*}

\subsection{3D MIDAS dust coverage maps}\label{appendix_3D_dust_coverage_maps}

\begin{figure*}
\includegraphics[width=\textwidth, height=10cm]{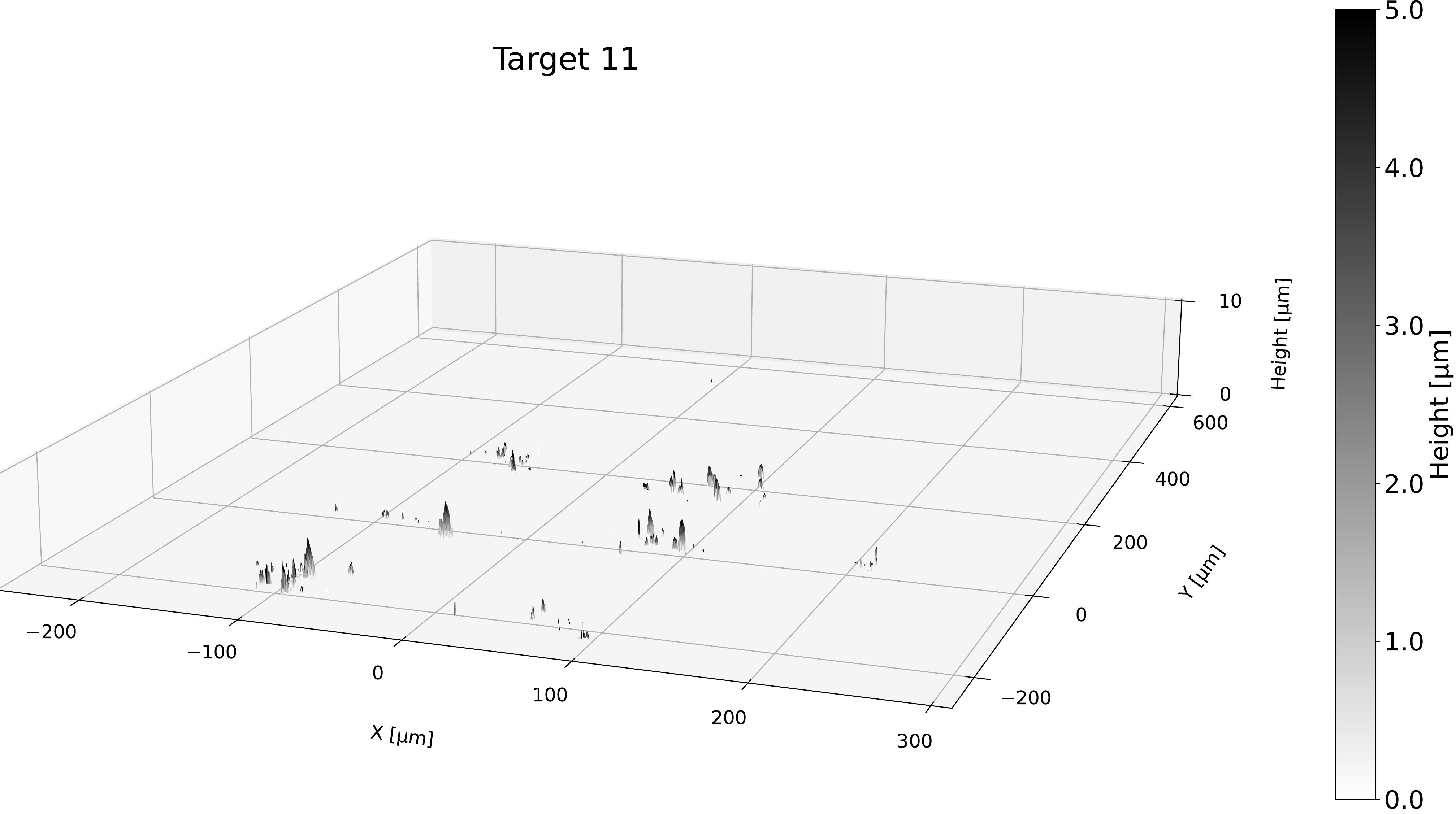}
\caption{3D dust coverage map of target 11 including only fully imaged particles as selected in Section~\ref{sec:results_Basic_statistics}.}
\label{fig:3D_dust_coverage_map_11}
\end{figure*}

\begin{figure*}
\includegraphics[width=\textwidth, height=10cm]{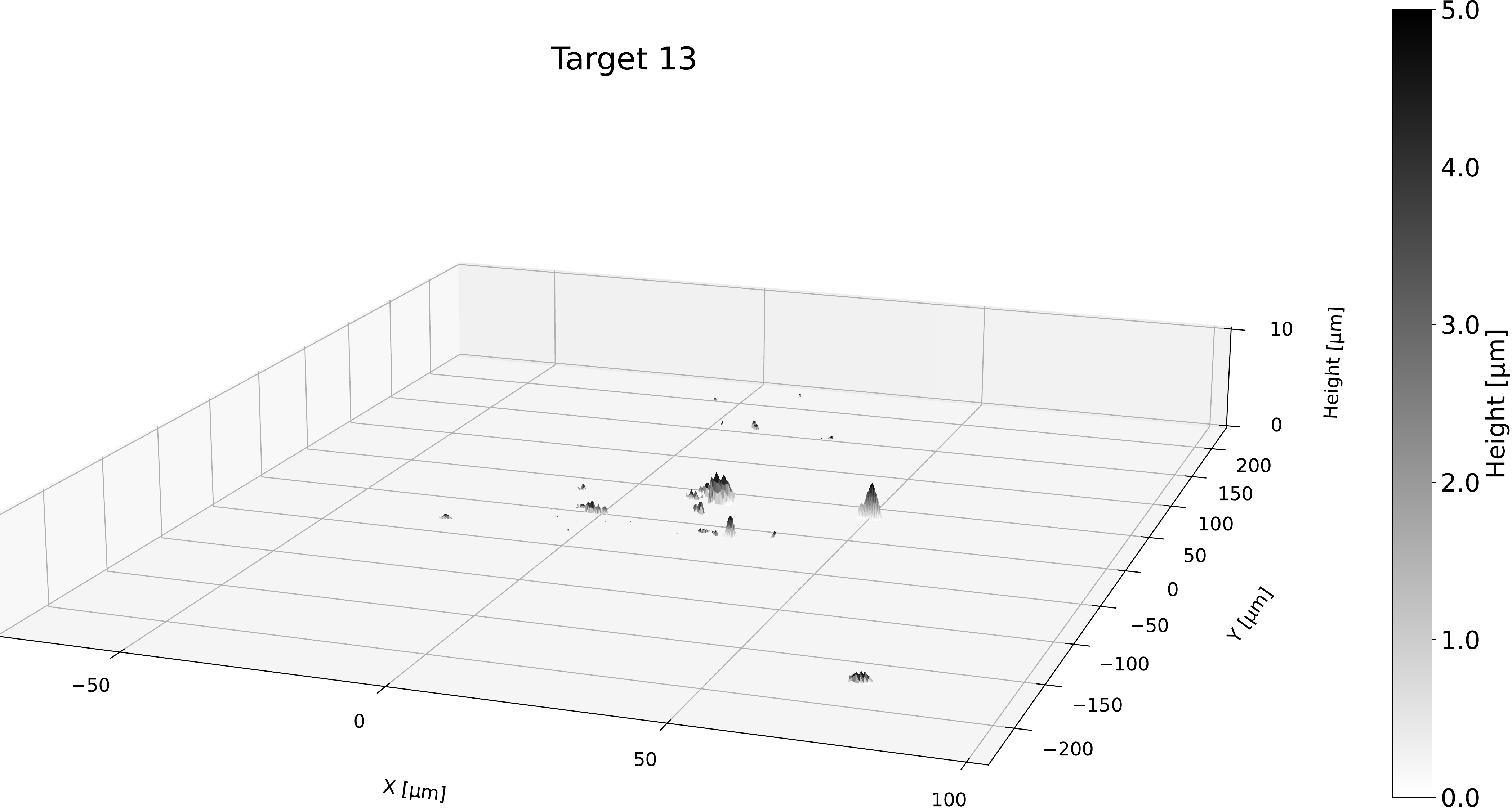}
\caption{3D dust coverage map of target 13 including only fully imaged particles as selected in Section~\ref{sec:results_Basic_statistics}.}
\label{fig:3D_dust_coverage_map_13}
\end{figure*}

\begin{figure*}
\includegraphics[width=\textwidth, height=10cm]{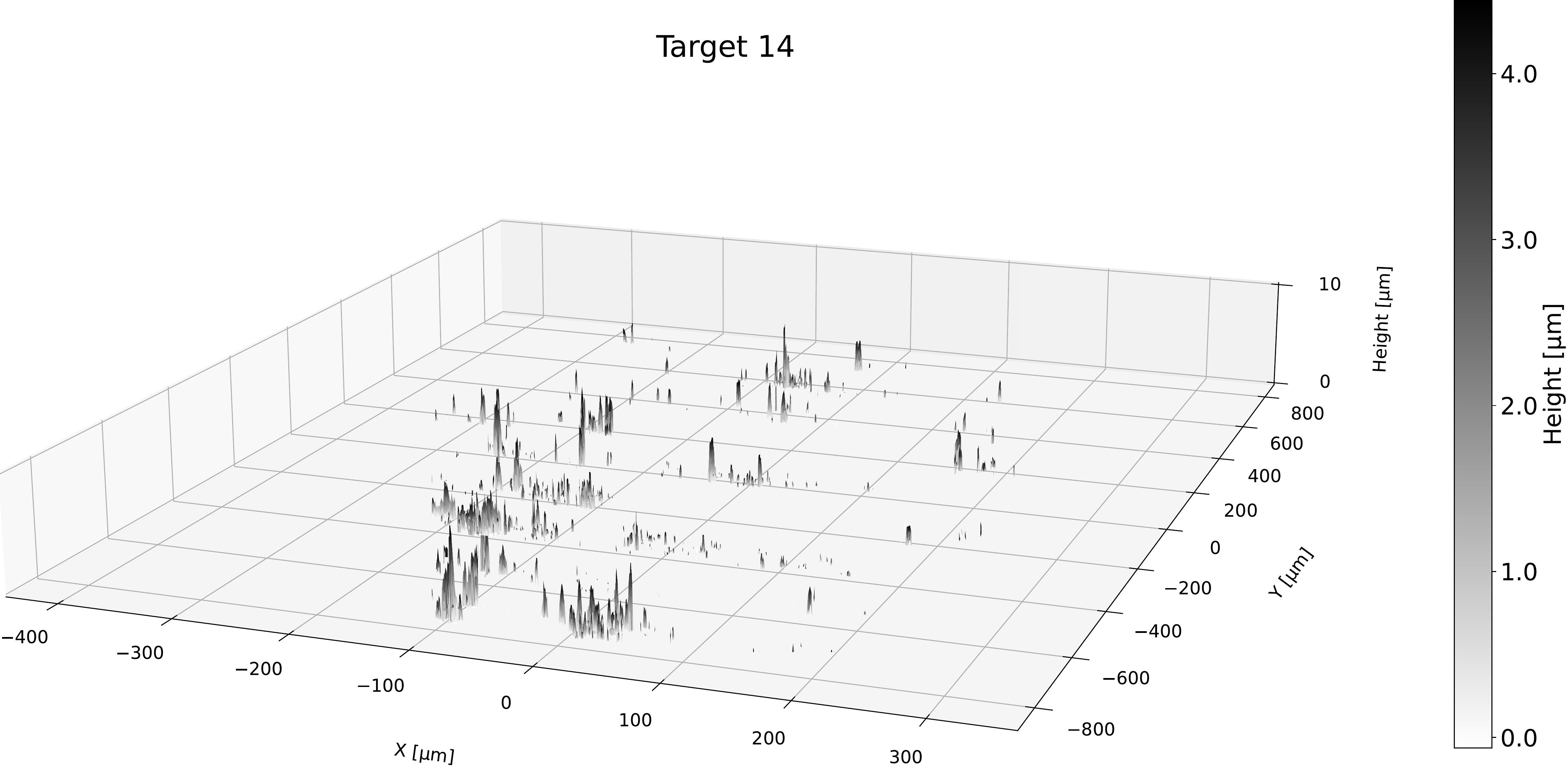}
\caption{3D dust coverage map of target 14 including only fully imaged particles as selected in Section~\ref{sec:results_Basic_statistics}.}
\label{fig:3D_dust_coverage_map_14}
\end{figure*}

\begin{figure*}
\includegraphics[width=\textwidth, height=10cm]{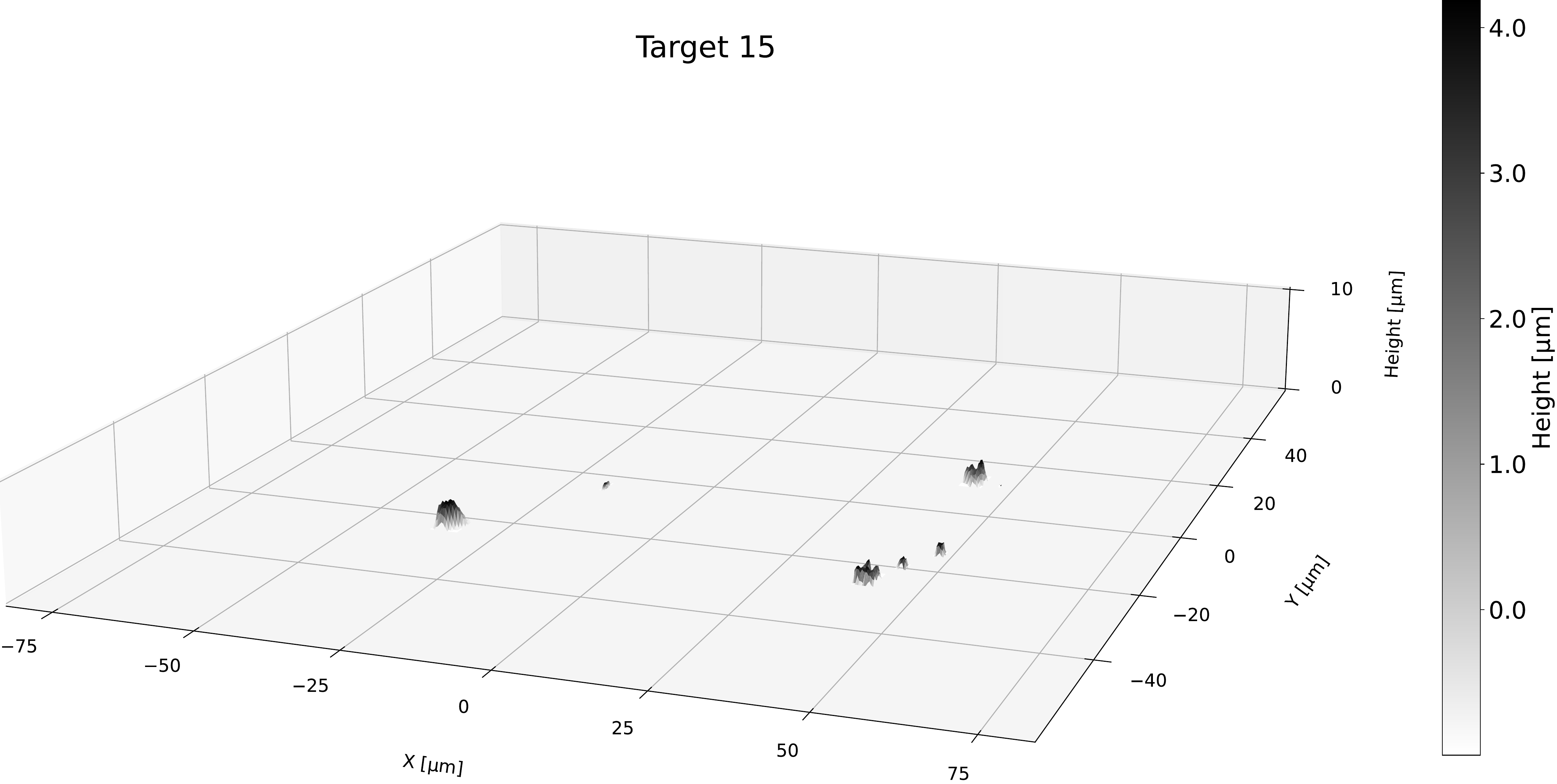}
\caption{3D dust coverage map of target 15 including only fully imaged particles as selected in Section~\ref{sec:results_Basic_statistics}.}
\label{fig:3D_dust_coverage_map_15}
\end{figure*}

\section{Visualization of the MIDAS dust clustering maps}\label{appendix_dust_clustering_maps}

The MIDAS dust clustering maps were created based on the MIDAS dust coverage maps (see Section~\ref{sec:The_size_of_MIDAS_clusters}) after running the mean shift clustering algorithm as described in Section~\ref{sec:clustering algorithm} and~\ref{sec:results_Fragmentation}. 
The maps are generally 2D images of the MIDAS targets with the 2D projection of the dust particles color-coded after their affiliation to a cluster. The cluster center is indicated by a semi-transparent red circle. The circle size does not indicate the cluster size as it is the same size for each cluster, namely, it only indicates the centroid (i.e., the center of mass with uniform density). For large clusters, some fragments can lie outside the red circle. Particles with the same color belong to the same cluster (the same parent particle) on the respective target. 

\begin{figure*}
\centering
\includegraphics[width=18cm, height=21cm]{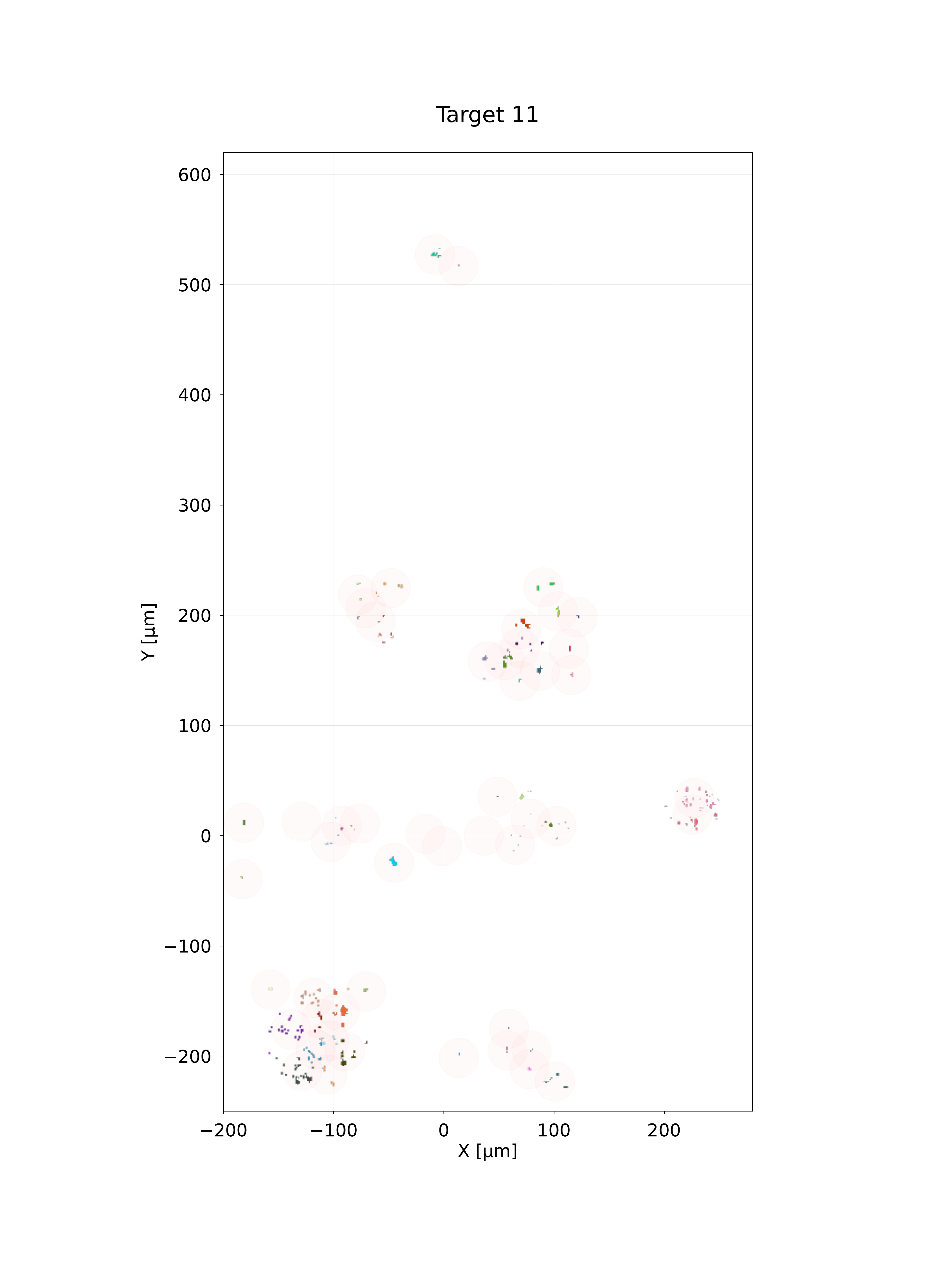}
\caption{Particle clustering map of target~11. Particles in the same cluster are depicted in the same color and red circles indicate the approximated location of the clusters.}
\label{fig:2D_dust_clustering_map_10}
\end{figure*}

\begin{figure*}
\centering
\includegraphics[width=17cm, height=20cm]{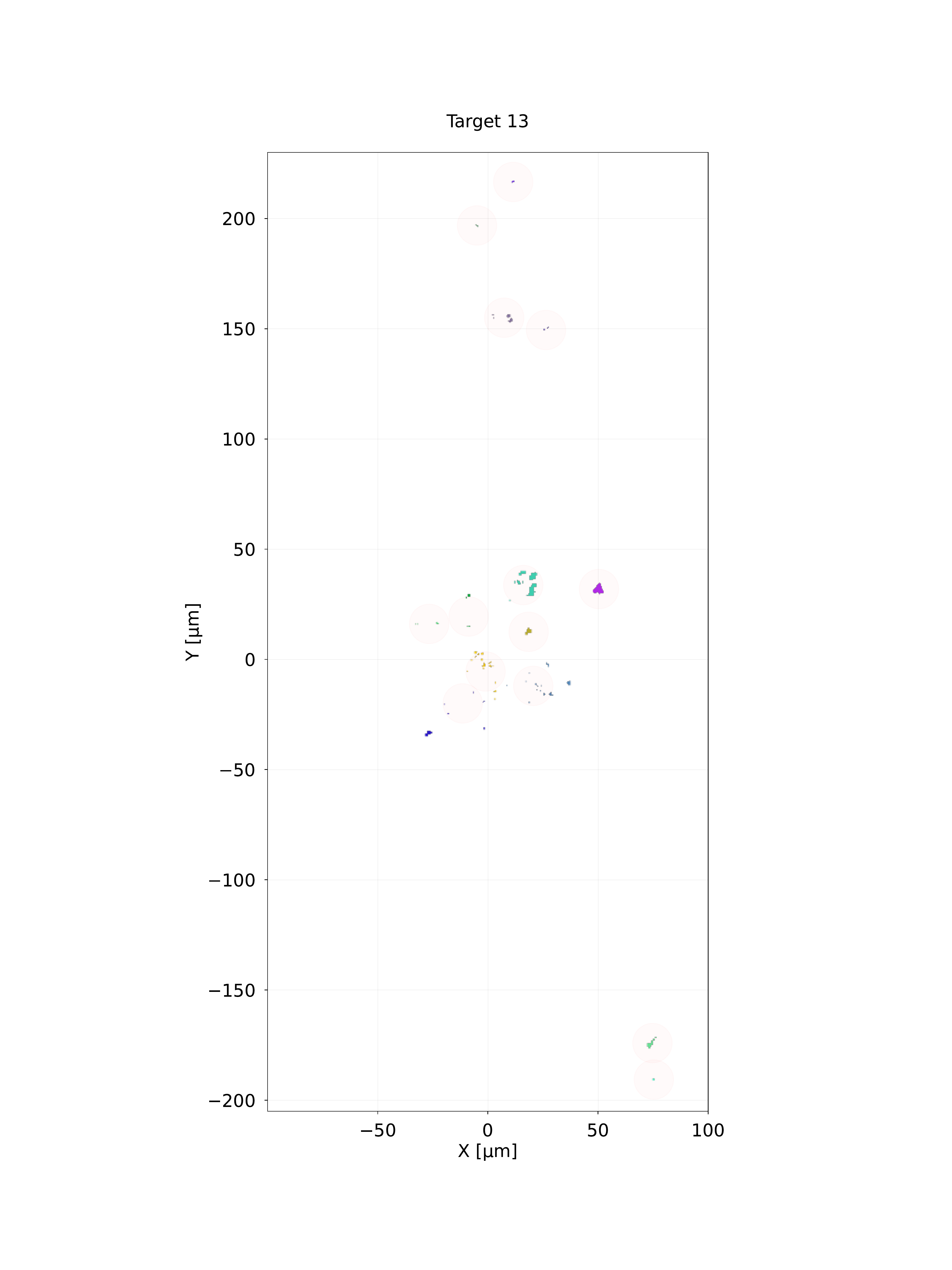}
\caption{Particle clustering map of target 13. Particles in the same cluster are depicted in the same color and red circles indicate the approximated location of the clusters.}
\label{fig:2D_dust_clustering_map_12 and 14}
\end{figure*}

\begin{figure*}
\centering
\includegraphics[width=20cm, height=24cm]{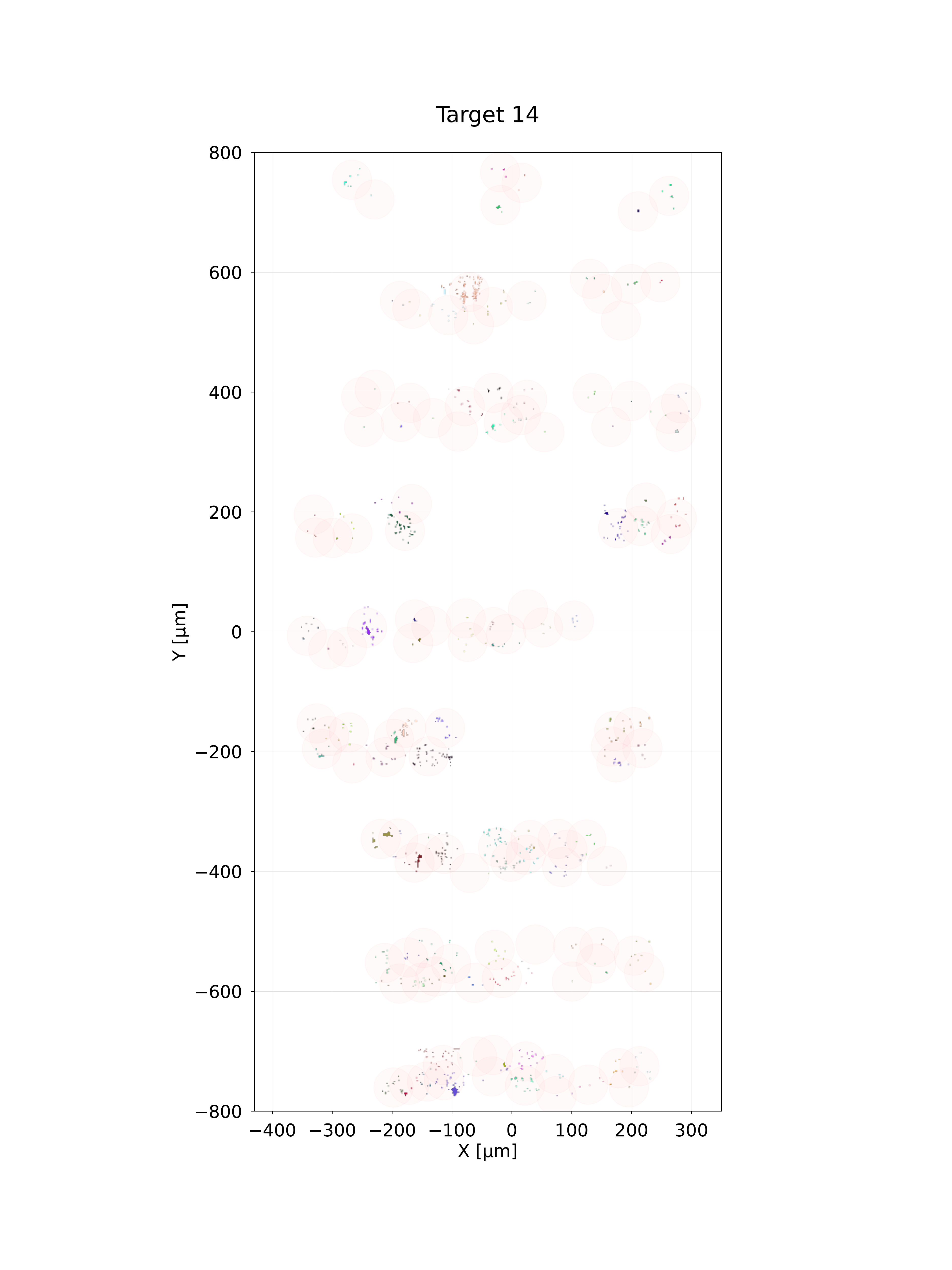}
\caption{Particle clustering map of target 14.  Particles in the same cluster are depicted in the same color and red circles indicate the approximated location of the clusters.}
\label{fig:2D_dust_clustering_map_13}
\end{figure*}

\begin{figure*}
\centering
\includegraphics[width=15 cm, height=18cm]{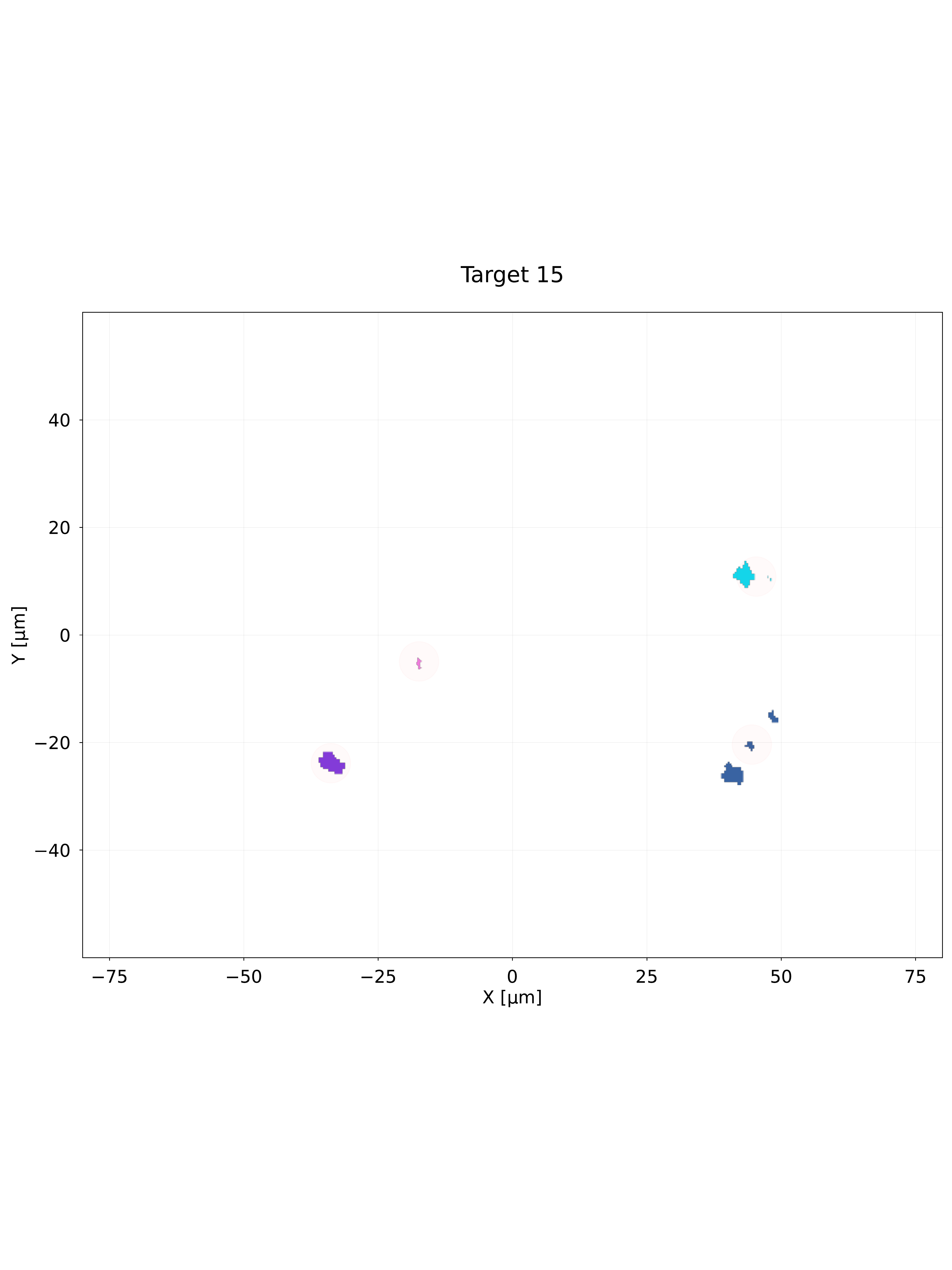}
\caption{Particle clustering map of target 15.  Particles in the same cluster are depicted with the same color and red circles indicate the approximated location of the clusters.}
\label{fig:2D_dust_clustering_map_14}
\end{figure*}

\end{appendix}
        
\end{document}